\documentclass[%
reprint,
amsmath,amssymb,
aps,
pra,
floatfix,
]{revtex4-2}

\usepackage{amsmath}
\usepackage{bm}
\usepackage{graphicx}
\usepackage{amssymb}
\usepackage{times}
\usepackage{color}
\usepackage{multirow}
\usepackage{braket}
\usepackage{hyperref}
\usepackage{chemformula}
\usepackage{makecell}
\usepackage{xfrac}
\usepackage{threeparttable}

\usepackage{graphicx}
\usepackage{subcaption}
\usepackage{ifthen}
\usepackage{xifthen}
\usepackage{multirow}
\usepackage{booktabs}
\usepackage{xcolor}
\usepackage[percent]{overpic}

\usepackage[
    protrusion=true,
    activate={true,nocompatibility},
    final,
    tracking=true,
    kerning=true,
    spacing=true,
    factor=1100]{microtype}
\microtypecontext{spacing=nonfrench}
\SetTracking{encoding={*}, shape=sc}{40}

\definecolor{optimumcolor}{HTML}{d62728}
\definecolor{mplblue}{HTML}{1f77b4}
\definecolor{mplorange}{HTML}{ff7f0e}
\definecolor{mplgreen}{HTML}{2ca02c}

\begin{document}

\title{Prescreening Point Defects in Semiconductors With Machine Learning}

\author{Paul Karlsson}
\affiliation{Department of Physics, Chemistry and Biology, Link\"oping University, Link\"oping, Sweden}

\author{Joel Davidsson}
\email{joel.davidsson@liu.se}
\affiliation{Department of Physics, Chemistry and Biology, Link\"oping University, Link\"oping, Sweden}

\author{Rickard Armiento}
\email{rickard.armiento@liu.se}
\affiliation{Department of Physics, Chemistry and Biology, Link\"oping University, Link\"oping, Sweden}

\begin{abstract}
High-throughput calculations using density-functional theory (DFT) are commonly used to explore point defects for applications in power electronics and quantum technologies. There is currently a major shift away from these traditional simulation techniques towards machine learning (ML) methods. We explore a class of physics-guided ML models for predicting defect formation energies and zero-phonon lines (ZPL) to identify point defects for quantum applications. The models are specifically targeted for use in a prescreening step for accelerated high-throughput workflows, and are therefore designed to avoid the costly relaxation step typically present with ML interatomic potentials (MLIPs). We compare performance for single and double point defect systems in 4H-SiC with ridge, kernel ridge, and multilayer perceptron (MLP) models using three different descriptors representing the defect systems. For vacancies and substitutions, the optimized models give mean absolute errors (MAEs) of 0.478 eV for the formation energy and 0.202 eV for ZPLs, which is just above the level at which such predictions can be useful even beyond the targeted prescreening, i.e., in some applications they may completely replace the need for costly DFT calculations. For interstitials the MAEs are larger, 1.123 eV for the formation energy and 0.230 eV for the ZPL, which, while still useful for prescreening, will not generally be useful for more detailed characterization. Hence, while the results may be further improved by model design and optimization, the models presented in this work are already useful for prescreening in high-throughput characterization of point defects.

\end{abstract}

\maketitle

\section{Introduction}

Silicon carbide is extensively used as a semiconductor in power electronics due to its relevant combination of properties, such as its wide band gap, high breakdown voltage, and high thermal conductivity \cite{current_sic_technology_for_power_electronic_devices_beyond_si,silicon_carbide_as_a_platform_for_power_electronics}.
Important applications range from motor controls, power supplies, robotics, transportation, renewable energy, telecommunication, and heating to power transmission.
Among the more than 200 polytypes \cite{silicon_carbide_fundamentals}, 4H-SiC is the most widely used \cite{joel_phd} and it already benefits from industrial production methods that keep improving \cite{silicon_carbide_fundamentals, development_of_n_type_epitaxial_growth_on_200_mm_4h_sic_wafers_for_the_next_generation_of_power_devices,physical_vapor_transport_growth_of_4h_silicon_carbide_single_crystals_by_a_tiling_method,recent_advances_in_4h_sic_epitaxy_for_high_voltage_power_devices,advances_in_fast_4h_sic_crystal_growth_and_defect_reduction_by_high_temperature_gas_source_method,a_new_model_for_in_situ_nitrogen_incorporation_into_4h_sic_during_epitaxy,bulk_and_epitaxial_growth_of_silicon_carbide,silicon_carbide_as_a_platform_for_power_electronics}. It is important to identify and control point defects in 4H-SiC semiconductors, as they can affect the carrier lifetime, which in turn can negatively impact conductivity, lead to excessive switching loss, and limit the switching frequency \cite{investigation_of_carrier_lifetime_in_4h_sic_epilayers_and_lifetime_control_by_electron_irradiation,Hazdra2021}.
At the same time, point defects in 4H-SiC have emerged as promising systems for applications in quantum computation, communication, sensing, and metrology \cite{diamon_and_silicon_converge,quantum_computation_and_quantum_information,quantum_computers,sic_color_centers_quantum_applications}. Quantum computers are of special interest since they may be able to compromise certain classical cryptographic protocols \cite{Bernstein2017}, enable quantum machine learning \cite{quantum_machine_learning}, and efficiently simulate quantum systems \cite{quantum_computation_and_quantum_information}. These simulations can then be used to develop novel artificial nanotechnologies and to further understand biological molecules \cite{quantum_computers}.

Machine learning (ML) has the potential to substantially reduce the computational expense of high-throughput calculations \cite{dft_expensive_1,dft_expensive_2} and is therefore attractive for exploring the large chemical and configurational space of possibly interesting point defects \cite{quantum_defects_by_design,adaq,ml_previous_10}. Prior works have used ML to predict formation energies and related stability measures for materials \cite{ml_material_previous_2,ml_material_previous_3,ml_material_previous_5} and point defects \cite{ml_previous_6,ml_previous_13,ml_previous_14,ml_previous_15,ml_previous_16,ml_previous_17,ml_previous_18,ml_previous_19,ml_previous_20,ml_previous_21,ml_previous_22,ml_previous_23,ml_previous_24,ml_previous_25,kumagai_oxygen_vacancies_ml,witman_defect_gnn}. Related work has addressed defect-structure exploration, including ML-accelerated structural reconstructions \cite{ml_previous_9} and systematic approaches for identifying defect ground-state structures \cite{mosquera_lois_defect_ground_states}, while ML methods have also been used to accelerate vibrational and thermal-property calculations \cite{ml_previous_11,ml_previous_12} and optical-property calculations \cite{ml_previous_8,turiansky_ml_optical_lineshapes}. Recent reviews summarize the growing use of data-driven and ML methods for point-defect simulations \cite{ml_previous_7,kumagai_kiyohara_ml_defects_review}. However, comparatively less attention has been given to the prediction of formation energies of interstitial defects, and to the prediction of zero-phonon lines (ZPLs) for vacancies, substitutions, and interstitials. These quantities are important for identifying point defects and assessing the potential of the point defects for quantum technology \cite{adaq}. Furthermore, available ML models do not yet come with the same track record of proven reliable predictions as density-functional theory (DFT) calculations, nor are models available for all properties of interest when characterizing a defect for possible use in quantum applications. Hence, our primary interest in this work is to design ML methods suitable for very broad \emph{prescreening} of an enumeration of candidate systems. The models of interest are thus meant to incur minimal computational cost per system during the screening phase, preferably avoiding expensive iterative procedures such as the relaxations typically present for ML predictions using interatomic potentials (MLIPs). Defect systems identified during prescreening are then meant to be followed up and validated by more computationally demanding but established calculations, including determining their relaxed atomic configurations.

In this paper, we investigate the ability of ridge, kernel ridge, and multilayer perceptron models to predict both formation energies and ZPL values for all three types of defects. The mean absolute error is minimized using 10-fold cross-validation, where we seek to come as close as possible to the error of high-throughput calculations used for screening defects, which can be estimated to around \(0.2\) eV for ZPLs and \(0.4\) eV for formation energies (see Sec.~\ref{sec:adaq}). The models are trained on a dataset of \(39333\) unique double and \(853\) unique single point defects in 4H-SiC, consisting of vacancies, substitutions, and interstitials with various charge and spin configurations that are taken from the database of ADAQ (Automatic Defect Analysis and Qualification) \cite{adaq,joel_phd,adaq_sic_1, adaq_sic_2}. ADAQ is a computational framework for point defects in semiconductors with automated workflows for high-throughput calculations of formation energies and magneto-optical properties such as ZPLs. It is implemented using the high-throughput toolkit (\emph{httk}) \cite{ml_material_previous_4} using the Vienna Ab initio Simulation Package (VASP) \cite{vasp1,vasp2} for density functional theory (DFT) calculations \cite{dft1,dft2,dft3}. We design and evaluate three descriptors that can be constructed at very low computational cost from an abstract defect representation, i.e., without requiring the full relaxed atomic coordinates: the Atomic Descriptor, the Wyckoff Letter Descriptor, and the Wyckoff Coordinate Descriptor. The first descriptor does not use any information about the symmetries of the atomic structure surrounding the point defects, whereas the last two do. The common input parameters for all three are: an atomic number representation for each point defect type, the distance between all unique pairs of point defects, and lastly, the charge and spin of the defect cluster.

The outline of the paper is as follows: the background covers point defects in 4H-SiC (\ref{sec:point_defects_in_4h_sic}), statistical measures (\ref{sec:statistical_measures}), ML models (\ref{sec:machine_learning_models}), and data scaling (\ref{sec:data_scaling}). The method section presents the descriptors (\ref{sec:descriptors}), details on the performance assessments (\ref{sec:performance_assessment}), implementation and optimization of the ML models (\ref{sec:implementation_and_optimization}). Section~\ref{sec:data_sets} presents the datasets used in this work.
Our results are presented in Sec.~\ref{sec:results} with all final scores and corresponding ML hyperparameters, followed by discussion (\ref{sec:discussion}) and conclusions (\ref{sec:conclusion}).

\section{Background}\label{sec:background}

\subsection{Point Defects in 4H-SiC}\label{sec:point_defects_in_4h_sic}

The three main types of point defects are vacancies, substitutions, and interstitials. Vacancies are created by removing an atom, substitutions are made by substituting an atom with another, and interstitials are formed by adding an atom at a previously unoccupied position not belonging to the original lattice. Groups of multiple point defects are called defect clusters \cite{joel_phd}.

There are different ways to describe the atomic structure symmetry surrounding point defects, which naturally depend on the crystal structure. In Ramsdell's notation, the polytype name 4H-SiC comes from its four atom layers in the unit cell (ABCB stacking) and its hexagonal crystal system (H). Locally, the atoms surrounding each lattice site in this polytype can either have a hexagonal structure, denoted \emph{h}, or a cubic structure, denoted \emph{k} \cite{silicon_carbide_fundamentals}. Point defects, namely vacancies and substitutions, at these two sites normally exhibit distinct optical and electronic signatures. If the actual properties vary accordingly, then it is useful in quantum technology applications where it could enable precise selection of coherence properties and emission wavelength \cite{color_centers_in_silicon_carbide}. Local site symmetry of interstitials are instead described using Wyckoff positions. These positions range from points, lines, and planes to the entire space \cite{bilbao_crystallographic_server_1,bilbao_crystallographic_server_2,bilbao_crystallographic_server_3}. It is related to site symmetry, which refers to how the structure symmetry of neighboring atoms remains the same under symmetry operations such as rotations \cite{International_tables_wyckoff}.

The formation energy, the energy needed to create a point defect, is a marker of stability \cite{adaq} given by
\begin{equation}
\Delta H_{D,q}(E_f, \mu) = \left[ E_{D,q} - E_H \right] + \sum_i n_i \mu_i + q E_f + E_{\mathrm{corr}}(q),
\label{eq:fe_energy}
\end{equation}
where \(E_f\) is the Fermi energy, \(E_{D,q}\) and \(E_H\) are respectively the total energies of the charged defect supercell and the host material supercell, \(i\) is the atomic number notation used in \(n_i\), which tracks the number of added or removed atoms, and \(\mu_i\), which represents the chemical potential, \(q\) is the charge of the defect supercell, and lastly, \(E_{\mathrm{corr}}\) corrects finite-size effects due to charge \cite{a_computational_framework_for_automation_of_point_defect_calculations,first_principles_calculations_for_point_defects_in_solids}.

Zero-phonon lines are spectroscopic features induced by point defects with states in the band gap that allow electrons to transition between. If the non-radiative decay rate is small, energy can be released as photoluminescence following an allowed transition between defect states. If a phonon is emitted or absorbed during the transition, the energy released is referred to as the emission or absorption phonon sideband. However, without the involvement of phonons, a more concentrated energy release occurs between the emission and absorption sidebands, and this is referred to as the zero-phonon line. The zero-phonon lines are useful for identifying point defects \cite{adaq, joel_phd} and for assessing their potential applications, such as single-photon emitters for quantum computers, quantum communication, and quantum metrology \cite{solid_state_single_photon_emitters}. The ZPL energy is given by
\begin{equation}
E_{\mathrm{ZPL}} = E_{e,\min} - E_{g,\min},
\label{eq:zpl_energy}
\end{equation}
where \(E_{e,\min}\) and \(E_{g,\min}\) are the energy minima of the excited state and the ground state, respectively \cite{adaq, joel_phd}.

\subsection{Statistical Measures}\label{sec:statistical_measures}

Our statistical analysis uses the standard deviation and covariance,
\begin{equation}
s = \sqrt{\frac{1}{n-1} \sum_{i=1}^{n} (x_i - \bar{x})^2},
\label{eq:standard_deviation}
\end{equation}
\begin{equation}
\mathrm{Cov}(x,y) = \frac{1}{n-1}\sum_{i=1}^{n}(x_i - \bar{x})(y_i - \bar{y})
\label{eq:covariance}
\end{equation}
where \(n\) is the number of data points of each variable, \(x_i, y_i\) are numeric values of the input parameter of data point \(i \in \{1,2,\ldots,n\}\),
and \(\bar{x}\) and \(\bar{y}\) are the averages of the two variables.
The covariance measures how much two variables \(x\) and \(y\) tend to increase and decrease together relative to their mean.

The Pearson correlation coefficient is a statistical measure of how linearly correlated two variables \(x\) and \(y\) are, given by
\begin{equation}
r_{xy} = \frac{\text{cov}(x, y)}{s_x s_y},
\label{eq:pearson_correlation_coefficient}
\end{equation}
where \(\text{cov}(x, y)\) is the covariance between \(x\) and \(y\), and \(s_x\) and \(s_y\) are the standard deviations of the variables \cite{pearson_correlation_coefficient}. Values of $r_{xy}$ between $-0.3$ and $0$; and between $0$ and $0.3$ indicate a weak negative or positive linear correlation respectively between \(x\) and \(y\). A value of \(0\) means there is no linear correlation and a value of 1 means there is a perfect positive correlation \cite{pearson_correlation_coefficient}.

The primary performance measure used to evaluate the performance of the ML model predictions in this work is the mean absolute error,
\begin{equation}
    \text{MAE} = \frac{1}{n} \sum_{i=1}^{n} \left| y_i - \hat{y}_i \right|,
    \label{eq:mean_absolute_error}
\end{equation}
where \(y_i\) and \(\hat{y}_i\) are correspondingly the true and the predicted value of data point \(i \in \{1,2,\ldots,n\}\).

\subsection{ML Models}\label{sec:machine_learning_models}

Ridge regression fits a linear function to the input parameters. Its prediction function \(f\) for centered inputs is
\begin{equation}
\begin{aligned}
f(\boldsymbol{x}) = \big( \underbrace{ \left(\boldsymbol{X}^{\mathrm{T}} \boldsymbol{X} + \alpha \boldsymbol{I}\right)^{-1} \boldsymbol{X}^{\mathrm{T}} \boldsymbol{y} }_{\boldsymbol{w}} \big)^\top \boldsymbol{x} + b,
\end{aligned}
\label{eq:ridge}
\end{equation}
where \(\boldsymbol{X}\) and \(\boldsymbol{y}\) contain all the input parameters and the true values of the training data, respectively, \(\alpha \geq 0\) is a regularization term, \(\boldsymbol{w}\) are the weights, \(\boldsymbol{x}\) is the input parameters, and \(b\) is the bias set to the average of \(\boldsymbol{y}\)  \cite{elements_statistical_learning}.

Kernel ridge regression transforms the input parameters so that an originally nonlinear function can be learned as a linear function \cite{foundation_ml}. In this paper, the transformation is carried out by the Gaussian kernel \cite{sklearn_rbf,elements_statistical_learning} and the prediction function \(f\) are
\begin{equation}
\begin{aligned}
    f(\boldsymbol{x}) &= \sum_{i=1}^{n} \lambda_i \underbrace{ e^{-\gamma \|\boldsymbol{x}_{\scriptstyle i} - \boldsymbol{x}\|^{\scriptstyle 2}} }_{\text{Gaussian Kernel}}, \\
    \boldsymbol{\lambda} &= \left( \boldsymbol{K} + \alpha \boldsymbol{I} \right)^{-1} \boldsymbol{y}, \\
    K_{ij} &= \exp(-\gamma||\boldsymbol{x_i}-\boldsymbol{x}_j||^2),
\end{aligned}
\label{eq:kernel_ridge}
\end{equation}
where \(\boldsymbol{x_i}\) and \(\lambda_i\) respectively represent the input parameters and the attributed weight of data point \(i \in \{1, \ldots, n\}\) in the training data, \(\gamma > 0\) controls how similarities (the Euclidean distance) between the input parameters of each data point in the training data and the point to be predicted affect the prediction, \(\alpha > 0\) is a regularization term, \(\boldsymbol{K}\) is the kernel matrix of the training data, and finally, \(\boldsymbol{y}\) contains the true values of the training data \cite{foundation_ml}. The loss function of the kernel ridge model is convex, which makes it easier to optimize \cite{the_loss_surface_of_multilayer_netweorks}.

The multilayer perceptron model is a type of artificial neural network. It is composed of fully connected layers consisting of processing units called neurons, which compute affine functions and pass the results to an activation function, before passing them forward to the next layer.
The four activation functions considered are: the identity function \(x\), the logistic function \(\frac{1}{1 + e^{{\scriptstyle -x}}}\), the hyperbolic tangent \(\tanh(x)\), and the ReLU function \(\max(0, x)\).

All weights of the affine functions are initialized randomly using a set seed referred to as the random state \cite{sklearn_mlpregressor}. Each weight is then updated iteratively by optimization algorithms based on how much it contributes to the prediction error (backpropagation). The contribution is calculated using the chain rule together with the gradient of the loss function \cite{deep_learning_python},
\begin{equation}
    \mathcal{L}(\boldsymbol{\hat{y}}, \boldsymbol{y}, \boldsymbol{x}) = \frac{1}{2n} \sum_{i=1}^{n} \|\hat{y}_i(\boldsymbol{x}) - y_i\|_2^2 + \frac{\alpha}{2n} \|\boldsymbol{W}\|_2^2,
    \label{eq:neural_loss}
\end{equation}
where \(\boldsymbol{\hat{y}}\) contains the predicted values, \(\boldsymbol{y}\) contains the true values, \(\boldsymbol{x}\) contains the input parameters of the prediction, \(n\) is the number of data points, \(\alpha\) is a regularization parameter, and \(\boldsymbol{W}\) contains all the weights \cite{sklearn_neural_network}. A drawback of the multilayer perceptron model is that the loss function is non-convex, meaning it is harder to optimize \cite{non_convex_optimization_method_for_machine_learning, sklearn_neural_network,the_loss_surface_of_multilayer_netweorks}.

The three employed optimization algorithms are: Nesterov Accelerated Gradient (NAG) \cite{nag}, ADAptive Moment estimation (Adam) \cite{adam}, and Limited memory Broyden-Fletcher-Goldfarb-Shanno (L-BFGS) \cite{lbfgs}. The weights are updated until a certain maximum number of iterations has been reached, L-BFGS has reached a maximum number of loss function evaluations, or the error does not improve more than the tolerance value after \(10\) consecutive iterations \cite{sklearn_mlpregressor}. Further details regarding the optimization algorithms are found in Supplementary Materials~S3. As a final point, the input parameters are typically scaled to speed up the optimization algorithms, which is further motivated in Section~\ref{sec:data_scaling}.

\subsection{Data Scaling}\label{sec:data_scaling}

Scaling of the numerical values used in ML models can be crucial to achieve good performance. The two main techniques are standardization and normalization \cite{hand_on_machine_learning}. In this study, both the input parameters and the target values are scaled.

Standardization (also known as z-score normalization) scales the input parameters to have an average of zero and a unitary standard deviation, putting everything on the same scale while preserving the underlying distribution of the data. The formula for standardizing the data points is given by
\begin{equation}
    x_{\text{standardized}} = \frac{x - \mu}{\sigma},
    \label{eq:input_parameter_standardization}
\end{equation}
where \(x_{\text{standardized}}\) is the standardized data point, \(x\) is the data point being scaled, and \(\mu\) and \(\sigma\) are the mean and standard deviation of all the data points of the specific input parameter, respectively \cite{hand_on_machine_learning,shallow_and_the_deep}.

Normalization (also commonly referred to as min-max scaling) consists of shifting and scaling the input parameters to fit the values within the closed interval \([0,1]\). The associated formula is
\begin{equation}
x_{\text{normalized}} = \frac{x - \min(\boldsymbol{x})}{\max(\boldsymbol{x}) - \min(\boldsymbol{x})},
\label{eq:min_max_scaling}
\end{equation}
where \(x_{\text{normalized}}\) is the normalized data point, \(x\) is the data point getting scaled, and \(\boldsymbol{x}\) are all the data points for the specific input parameter \cite{hand_on_machine_learning,shallow_and_the_deep}.

Scaling the data through standardization and normalization is particularly suited for artificial neural networks. By centering the inputs and bringing them to comparable scales, standardization can reduce differences in the magnitudes of the corresponding synaptic weight updates \cite{haykin_neural} and has been found to improve the performance of artificial neural networks \cite{standardization_neural_1,standardization_neural_2,standardization_neural_3}.

\section{Method}\label{sec:method}

\subsection{Descriptors for defect systems}\label{sec:descriptors}

The representation that encodes the input for an ML model as parameters, known as the descriptor, can be crucial for model performance \cite{descriptor_definition, descriptor, ml_material_previous_3}. This work explores descriptors created from the following input parameters: an atomic number representation of each point defect, the distance in Ångström between each unique pair of point defects, the charge and spin state of the defect cluster, and optionally, the local site symmetries of the point defects. For vacancies, the atomic number representation consists of the negative atomic number of the removed atom. The positive atomic number corresponds to the added atom in substitutions and interstitials. A zero represents the absence of a point defect. Lastly, for single point defects, the distance is set to zero.

These particular input parameters are motivated from the physics of defect systems: the atomic number representation carries information about the point defect type; both the distance and the local site symmetries provide information about the point defect symmetry and its position relative to the general crystal structure; the charge state is associated with the spin state, and both tend to give information about optical and electronic properties of the point defect. Furthermore, the point defect symmetry is known to influence the zero-phonon lines \cite{silicon_carbide_novel_quantum_technology}.

In more detail, the first descriptor is the \emph{Atomic Descriptor}, which uses all input parameters except the local site symmetries, encoded as
\begin{equation}
\boldsymbol{x} =
\begin{bmatrix}
a_1 & \cdots & a_n & d_{12} & \cdots & d_{(n-1)n} & c & s
\end{bmatrix}^{\top},
\label{eq:descriptor_vector_no_symmetry}
\end{equation}
where \(n\) is the fixed number of allocated defect slots. The indices \(i,j \in \{1,2,\ldots,n\}\) refer to these slots. The entry \(a_i\) is the atomic number representation of each point defect assigned to slot \(i\), or zero if the slot is unused. For \(i<j\), \(d_{ij}\) is the distance between the point defects in slots \(i\) and \(j\), or zero if either slot is unused. Finally, \(c\) and \(s\) are the charge and spin state of the defect cluster, respectively. The slots are sorted by their atomic number representations, with the distance entries following the same ordering, to obtain a consistent representation of each defect cluster.

The second descriptor is the \emph{Wyckoff Letter Descriptor}. It extends the Atomic Descriptor by assigning fixed groups of entries in the descriptor vector to site categories and placing the atomic number representation of each defect in the group corresponding to its category. These local site symmetries are either hexagonal \emph{h} or cubic \emph{k} for vacancies and substitutions, and for interstitials they are represented by the Wyckoff letter of the position of the point defect according to Table~\ref{tab:wyckoff_letters_coordinates}. The descriptor then takes the form
\begin{equation}
\boldsymbol{x} =
\begin{bmatrix}
a_{1_{\scriptstyle 1}} & \dots & a_{n_{\scriptstyle m}} &
d_{1_{\scriptstyle 1}2_{\scriptstyle 1}} & \dots & d_{{n-1}_{\scriptstyle m}n_{\scriptstyle m}} &
c & s
\end{bmatrix}^\top
\label{eq:descriptor_vector_symmetry}
\end{equation}
where \(a_{i_{\scriptstyle k}}\) is the atomic number representation of each point defect assigned to defect slot \(i \in \{1,2,\ldots,n\}\) having the neighboring atomic structure symmetry \(k \in \{1,2,\ldots,m\}\), \(d_{i_{\scriptstyle u}j_{\scriptstyle v}}\) is the distance between point defect \(i\) and \(j\) respectively having the local site symmetries \(u\) and \(v\), and \(c\) and \(s\) are the charge and the spin state of the defect cluster. The atomic number representations are sorted by their value within each subgroup corresponding to a certain neighboring atomic structure symmetry, to further ensure a unique representation of each defect cluster.

However, this work only considers single and double defects, for which there is at most one non-zero distance. Hence, the descriptor is simplified to
\begin{equation}
\boldsymbol{x} =
\begin{bmatrix}
a_{1_1} & a_{2_1} & \cdots &
a_{1_m} & a_{2_m} & d & c & s
\end{bmatrix}^{\mathsf T},
\label{eq:descriptor_vector_symmetry_simple}
\end{equation}
with \(n=2\) slots allocated per category for both single and double defects with the unoccupied atomic-number slots zero, and the distance between the two point defects ($d$) is included only once, and set to zero for single defects.

\begin{table*}
\centering
\renewcommand{\arraystretch}{1.6}
\setlength{\tabcolsep}{6pt}
\centering
\small
\begin{tabular}{l l | l l | l l | l l}
\toprule
Letter & Coordinate & Letter & Coordinate & Letter & Coordinate & Letter & Coordinate \\
\midrule

a & $(0,0,\scalebox{1.15}{$\frac{3}{32}$})$ & b & $(\scalebox{1.15}{$\frac{1}{3}$},-\scalebox{1.15}{$\frac{1}{3}$},\scalebox{1.15}{$\frac{3}{32}$})$ & c & $(\scalebox{1.15}{$\frac{45}{82}$},-\scalebox{1.15}{$\frac{45}{82}$},\scalebox{1.15}{$\frac{8}{85}$})$ & c & $(\scalebox{1.15}{$\frac{37}{82}$},-\scalebox{1.15}{$\frac{37}{82}$},\scalebox{1.15}{$\frac{8}{85}$})$ \\
a & $(0,0,\scalebox{1.15}{$\frac{11}{32}$})$ & b & $(\scalebox{1.15}{$\frac{1}{3}$},-\scalebox{1.15}{$\frac{1}{3}$},\scalebox{1.15}{$\frac{11}{32}$})$ & c & $(\scalebox{1.15}{$\frac{5}{6}$},-\scalebox{1.15}{$\frac{5}{6}$},\scalebox{1.15}{$\frac{3}{32}$})$ & c & $(\scalebox{1.15}{$\frac{21}{88}$},-\scalebox{1.15}{$\frac{21}{88}$},\scalebox{1.15}{$\frac{7}{32}$})$ \\
b & $(\scalebox{1.15}{$\frac{1}{3}$},-\scalebox{1.15}{$\frac{1}{3}$},\scalebox{1.15}{$\frac{7}{32}$})$ & b & $(\scalebox{1.15}{$\frac{1}{3}$},-\scalebox{1.15}{$\frac{1}{3}$},\scalebox{1.15}{$\frac{19}{32}$})$ & c & $(\scalebox{1.15}{$\frac{7}{74}$},-\scalebox{1.15}{$\frac{7}{74}$},\scalebox{1.15}{$\frac{7}{32}$})$ & c & $(\scalebox{1.15}{$\frac{4}{7}$},-\scalebox{1.15}{$\frac{4}{7}$},\scalebox{1.15}{$\frac{7}{32}$})$ \\
b & $(\scalebox{1.15}{$\frac{1}{3}$},-\scalebox{1.15}{$\frac{1}{3}$},\scalebox{1.15}{$\frac{27}{32}$})$ & c & $(\scalebox{1.15}{$\frac{3}{7}$},-\scalebox{1.15}{$\frac{3}{7}$},\scalebox{1.15}{$\frac{7}{32}$})$ & c & $(\scalebox{1.15}{$\frac{55}{96}$},-\scalebox{1.15}{$\frac{55}{96}$},\scalebox{1.15}{$\frac{15}{32}$})$ & c & $(\scalebox{1.15}{$\frac{19}{99}$},-\scalebox{1.15}{$\frac{19}{99}$},\scalebox{1.15}{$\frac{3}{32}$})$ \\
b & $(\scalebox{1.15}{$\frac{1}{3}$},-\scalebox{1.15}{$\frac{1}{3}$},\scalebox{1.15}{$\frac{29}{32}$})$ & c & $(\scalebox{1.15}{$\frac{41}{96}$},-\scalebox{1.15}{$\frac{41}{96}$},\scalebox{1.15}{$\frac{15}{32}$})$ & c & $(\scalebox{1.15}{$\frac{51}{67}$},-\scalebox{1.15}{$\frac{51}{67}$},\scalebox{1.15}{$\frac{15}{32}$})$ & d & $(\scalebox{1.15}{$\frac{2}{7}$},\scalebox{1.15}{$\frac{2}{7}$},\scalebox{1.15}{$\frac{11}{32}$})$ \\
b & $(\scalebox{1.15}{$\frac{1}{3}$},-\scalebox{1.15}{$\frac{1}{3}$},\scalebox{1.15}{$\frac{31}{32}$})$ & c & $(\scalebox{1.15}{$\frac{48}{53}$},-\scalebox{1.15}{$\frac{48}{53}$},\scalebox{1.15}{$\frac{15}{32}$})$ & c & $(\scalebox{1.15}{$\frac{5}{6}$},-\scalebox{1.15}{$\frac{5}{6}$},\scalebox{1.15}{$\frac{7}{32}$})$ & & \\
b & $(\scalebox{1.15}{$\frac{1}{3}$},-\scalebox{1.15}{$\frac{1}{3}$},\scalebox{1.15}{$\frac{13}{32}$})$ & c & $(\scalebox{1.15}{$\frac{1}{2}$},-\scalebox{1.15}{$\frac{1}{2}$},\scalebox{1.15}{$\frac{13}{32}$})$ & c & $(\scalebox{1.15}{$\frac{1}{6}$},-\scalebox{1.15}{$\frac{1}{6}$},\scalebox{1.15}{$\frac{15}{32}$})$ & & \\[0.05cm]
\bottomrule
\end{tabular}
\caption{The Wyckoff letters and coordinates of the point defect sites of the 4H-SiC interstitial data \cite{adaq,bilbao_crystallographic_server_1,bilbao_crystallographic_server_2,bilbao_crystallographic_server_3}.}
\label{tab:wyckoff_letters_coordinates}
\end{table*}

The third descriptor is the \emph{Wyckoff Coordinate Descriptor}.
It takes the same mathematical form as the Wyckoff Letter Descriptor, eqn~\ref{eq:descriptor_vector_symmetry}, but with a different
set of site categories. The categories \(h\) and \(k\) for
vacancies and substitutions are kept, but each of the
26 distinct interstitial coordinate triples in
Table~\ref{tab:wyckoff_letters_coordinates} constitutes a separate category.
Hence, in eqn~\ref{eq:descriptor_vector_symmetry} the atomic number representations, pair distances, charge, and spin are unchanged, and \(a_{i_{\scriptstyle u}}\) remains the atomic number representation in slot \(i\) of category \(u\), or zero if that slot is unoccupied. However, for interstitials, \(u\) now identifies a particular listed coordinate triple, meaning that sites sharing a Wyckoff letter but having different listed coordinates occupy different groups of entries.
Furthermore, for the case of only single and double defects of this work, just as for the Wyckoff Letter Descriptor, the distance between the two point defects ($d$) is included only once, and set to zero for single defects, i.e., matching eqn~\ref{eq:descriptor_vector_symmetry_simple}.
More details on the descriptors are available in Supplementary Materials~S2.

\subsection{Performance Assessment}\label{sec:performance_assessment}

Ten-fold cross-validation is used to get an indication of the average performance independent of the data bias. Using a fixed seed, the double point defect data is first randomly split into 10 subgroups, called folds. The single point defect data is grouped with nine of the folds, which are then used for training the ML models. After the training is complete, the ML models try to predict the data points in the last remaining fold, called the test fold, and the accuracy of the predictions is measured using MAE. This process is repeated ten times for each unique fold combination and the average of the ten resulting scores is used for the final score \cite{hand_on_machine_learning,python_machine_learning}.
The MAE values are calculated separately for each test fold using predictions from a model trained on the corresponding training folds. The test data are excluded from model fitting, and no additional validation set is reserved outside the cross-validation procedure.

The learning curve shows how the performance varies with the amount of training data. These curves are obtained using a variation of 10-fold cross-validation, which is referred to as varied split ratio 10-fold cross-validation, described below. This variation varies the train-test split ratio nine times following eqn~\ref{eq:number_train_test_k_fold_cross_validation}, where \(N_{\text{Train}}\) and \(N_{\text{Test}}\) are, respectively, the numbers of folds used for training and testing out of the 10 original folds containing the data of the double point defects. The data of single point defects is once more always added to the training data. This process results in nine averaged mean absolute error scores which are plotted as a function of the mean number of training data points used for calculating each individual average score.

Additionally, the varied split ratio 10-fold cross-validation can be used to assess prediction performance on the training data.
If one assumes the training and validation learning curves converge near their midpoint as the training set grows, the mean of the fold-averaged training and validation MAEs at \(N_{\mathrm{Train}} = 9\) and \(N_{\mathrm{Test}} = 1\) can be taken as a rough estimate of the validation MAE in this limit:
\begin{equation}
\begin{aligned}
N_{\text{Train}} &\in \{1,2,\ldots,9\} \\
N_{\text{Test}} &= 10 - N_{\text{Train}}
\end{aligned}
\label{eq:number_train_test_k_fold_cross_validation}
\end{equation}

Lastly, scatter plots showing the predicted values as a function of the true values are yet another way to assess performance. If the predictions are accurate, then the points will line up on a straight line in the middle, whereas they will be more spread out if inaccurate.

\subsection{Implementation and Optimization}\label{sec:implementation_and_optimization}

The open-source library scikit-learn \cite{scikit-learn} is used for dataset scaling, the implementation of the ML models, the optimization, and performance assessment. The Ridge model is optimized by trying out the 100 different values of \(\alpha\) evaluated using 10-fold cross-validation with mean absolute error as the performance metric, given by
\begin{equation}
    \alpha \in \{10^{e/4}, \ e = -32, -31, \dots, 67 \}.
\label{eq:ridge_alpha_values}
\end{equation}

The kernel ridge model is optimized in the same way as the ridge model but with the following values
\begin{equation}
\begin{aligned}
\alpha &\in \{10^e, \ e=-8,-7,\dots,2 \}, \\
\gamma &\in \{0, \tfrac{1}{59}, \tfrac{1}{15}, \tfrac{1}{5}\} \,\cup\, \{10^e, \ e=-4,-3,\dots,4 \},
\end{aligned}
\label{eq:kernel_ridge_alpha_gamma_values}
\end{equation}
where the values of \(\gamma\) after the initial zero correspond to the inverse of the number of input parameters for the Wyckoff Coordinate Descriptor, the Wyckoff Letter Descriptor, and the Atomic Descriptor.

Optimizing the multilayer perceptron model consists once more of methodically trying out different hyperparameters. The hyperparameters that are kept constant throughout the whole optimization process and their values are the following: the maximum number of iterations and the maximum number of function calls for the L-BFGS algorithm are both set to infinite (in practice they have been set to \(9999999999999\)), the tolerance value is set to \(10^{-4}\), the learning rates in the NAG and Adam algorithms are set to \(10^{-3}\), \(\beta_1\), \(\beta_2\), and \(\epsilon\) for Adam are respectively set to \(0.9\), \(0.999\), and \(10^{-8}\) (standard values \cite{adam}), and lastly, the random state is set to 42. Additionally, the value of \(\alpha\) is kept constant at \(10^{-4}\) to begin with. The first step is to try different numbers of hidden layers with 25, 50, 75, and 100 neurons in each, together with all possible combinations of activation functions and optimization algorithms. When promising combinations are found they are then reiterated but with a different number of neurons, until a perceived optimum has been achieved. Up until this point only four out of the 10 possible scores in 10-fold cross-validations are used for the indicative average performance to save time and computational resources. Afterward, using all 10 scores in 10-fold cross-validation, a subset of the different values of \(\alpha\),
\begin{equation}
\alpha \in \{10^e, e = -12, -11, \ldots, 2 \}
\label{eq:neural_alpha}
\end{equation}
are tested for the most promising combinations until the optimization space has been explored satisfactorily. For the detailed optimization results of the models, please consult the supplementary material. 

\section{datasets}\label{sec:data_sets}

\subsection{ADAQ}\label{sec:adaq}

The datasets are extracted from the ADAQ defect database~\cite{adaq_database}.
The DFT results were produced for supercell (6x6x2, 576 atoms) with the semi-local functional by Perdew, Burke, and Ernzerhof (PBE)~\cite{PBE} as the exchange-correlation functional.
To enable high-throughput calculations, the computational settings for the screening workflow only use the gamma k-point, the electronic convergence criterion was set to $10^{-4}$~eV, and the ionic relaxation was considered converged when the energy change between successive ionic steps was below $5\times10^{-3}$~eV.
For additional computational details, see Ref.~\cite{adaq}.

Our use of settings appropriate for screening-level accuracy means that differences in MAEs for formation energy predictions below approximately 0.01 eV have no significance. In practice, the accuracy of the calculated formation energies are not well known due to the difficulty to precisely determine this quantity in experiments, e.g., Ayedh \textit{et al.} reports an experimental value for the formation energy of the carbon vacancy in 4H-SiC to be $5.0\pm0.2\ \mathrm{eV}$, where the error bars span values computed with accurate higher-order DFT methods \cite{Ayedh2014}. On that basis, one can argue that a reasonable estimate of the error of formation energy values calculated with screening-level DFT methods for 4H-SiC is given by the difference to the higher-order results, which is around $0.4~\mathrm{eV}$; however, across other methods and materials they can be larger, e.g., up to $1.0\ \mathrm{eV}$ \cite{Abbas2025}. Nevertheless, it is of course possible (and arguably meaningful for some applications) to train a model to reproduce the PBE DFT values with higher precision than their absolute accuracy.

There are more precise experiments of the ZPL, which allows us to estimate the errors in our screening-level DFT calculations in 4H-SiC to be around $0.2~\mathrm{eV}$ \cite{adaq, Abbas2025}. Furthermore, ADAQ only considers excitations with a vertical absorption energy above 0.4~eV at the ground state geometry. The structures for the selected excited states then undergo ionic relaxation, from which the ZPL is determined. Since this relaxation may result in a structure with a lower energy for the excited state than for the original ground state, some data points end up with a negative ZPL value. We investigated filtering these nonphysical results out during training and validation, but this did not significantly affect performance. Hence, we decided it was better to target faithfully predicting all the result from the ADAQ methodology, even for these results.

\subsection{Training and testing data}

The division of the data into different datasets depends on defect size (single or double defects), defect type (vacancies and substitutions or interstitials), and descriptor (Atomic, Wyckoff Letter, or Wyckoff Coordinate). The defect type referred to as interstitials means there is at least one interstitial in the defect cluster. In contrast, the joint defect type referred to as vacancies and substitutions includes only those types of defects.

Each dataset contains only one data point for each unique combination of input parameters, since the models cannot distinguish systems represented by identical inputs. If two or more data points share the same input parameters when creating the datasets, the data point with the lowest formation energy is kept. Consequently, the dataset used with the Atomic Descriptor has the fewest data points. To evaluate whether the removed data points affect the performance, two additional evaluations are made, labeled \emph{Wyckoff Letter Compare Descriptor} and \emph{Wyckoff Coordinate Compare Descriptor}. They refer to the application of the Wyckoff Letter Descriptor and the Wyckoff Coordinate Descriptor only to the data points used for the Atomic Descriptor, whose dataset is a strict subset of each of the two Wyckoff datasets. With these additional datasets, we arrive at a total of 20. For each dataset, the given names, the number of data points, the mean formation energy, and the mean zero-phonon line are given in Table~\ref{tab:dataset_details}.

\begin{table*}
\centering
\begin{tabular}{llcccccccc}
\toprule
& & \multicolumn{4}{c}{\textbf{Single}} & \multicolumn{4}{c}{\textbf{Double}} \\
\cmidrule(r){3-6} \cmidrule(r){7-10}
\textbf{Descriptor} & \textbf{Defect Type} & \textbf{Name} & \textbf{Nr. Points} & \textbf{Mean FE} & \textbf{Mean ZPL}
& \textbf{Name} & \textbf{Nr. Points} & \textbf{Mean FE} & \textbf{Mean ZPL} \\
\midrule
\multirow{2}{*}{Atomic}
& Vacancy-Substitution & A-VS-S & 170 & 5.86 & 0.1951 & A-VS-D & 5136 & 9.08 & 0.3189 \\
& Interstitial         & A-I-S  & 147 & 11.32 & 0.1780 & A-I-D  & 30502 & 10.20 & 0.2542 \\
\midrule
\multirow{2}{*}{W. Letter}
& Vacancy-Substitution & WL-VS-S & 306 & 5.81 & 0.1796 & WL-VS-D & 5834 & 9.14 & 0.3137 \\
& Interstitial         & WL-I-S  & 248 & 9.66 & 0.1767 & WL-I-D  & 32556 & 10.27 & 0.2549 \\
\midrule
\multirow{2}{*}{W. Coord.}
& Vacancy-Substitution & WC-VS-S & 306 & 5.81 & 0.1796 & WC-VS-D & 5834 & 9.14 & 0.3137 \\
& Interstitial         & WC-I-S  & 547 & 8.72 & 0.1380 & WC-I-D  & 33499 & 10.25 & 0.2537 \\
\midrule
\multirow{2}{*}{W. Letter C.}
& Vacancy-Substitution & WLC-VS-S & 170 & 5.86 & 0.1951 & WLC-VS-D & 5136 & 9.08 & 0.3189 \\
& Interstitial         & WLC-I-S  & 147 & 11.32 & 0.1780 & WLC-I-D  & 30502 & 10.20 & 0.2542 \\
\midrule
\multirow{2}{*}{W. Coord. C.}
& Vacancy-Substitution & WCC-VS-S & 170 & 5.86 & 0.1951 & WCC-VS-D & 5136 & 9.08 & 0.3189 \\
& Interstitial         & WCC-I-S  & 147 & 11.32 & 0.1780 & WCC-I-D  & 30502 & 10.20 & 0.2542 \\
\bottomrule
\end{tabular}
\caption{Names, number of data points, mean formation energy (FE) and mean zero-phonon line (ZPL) in eV for each dataset.}
\label{tab:dataset_details}
\end{table*}

\begin{figure}
    \centering

\makebox[\linewidth][c]{%
    \raisebox{-0.5\height}{\rotatebox{90}{\normalsize Value}}%
    \hspace{2mm}
    \begin{minipage}{0.9\linewidth}
    \centering

    \raggedleft
    \hspace{9mm}
    \begin{subfigure}[b]{0.98\linewidth}
        \centering
        \includegraphics[width=0.98\linewidth]{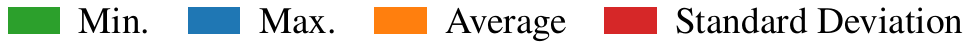}
    \end{subfigure}

    \centering
    \begin{subfigure}[b]{0.22\linewidth}
        \IfFileExists{figures/sic_no_wyckoff_vacancy-substitutional_input_parameter_details_unscaled_single.pdf}{
            \includegraphics[width=\linewidth]{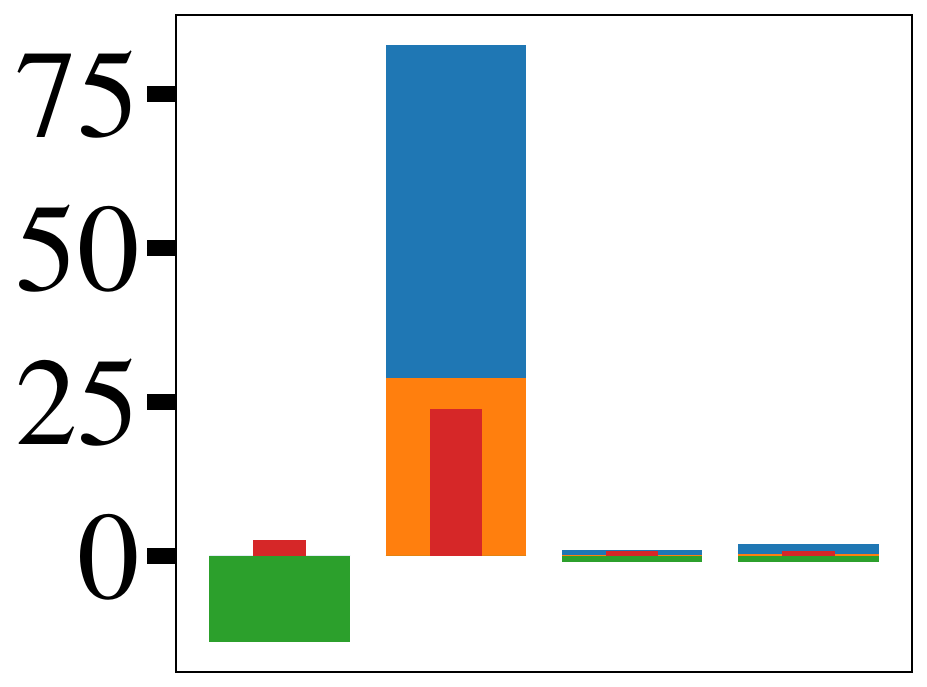}
        }{\centering Image not found}
        \vspace{-0.5cm}
        \caption{}\label{fig:input_parameter_details_a}
    \end{subfigure}
    \hfill
    \begin{subfigure}[b]{0.22\linewidth}
        \IfFileExists{figures/sic_no_wyckoff_vacancy-substitutional_input_parameter_details_unscaled_double.pdf}{
            \includegraphics[width=\linewidth]{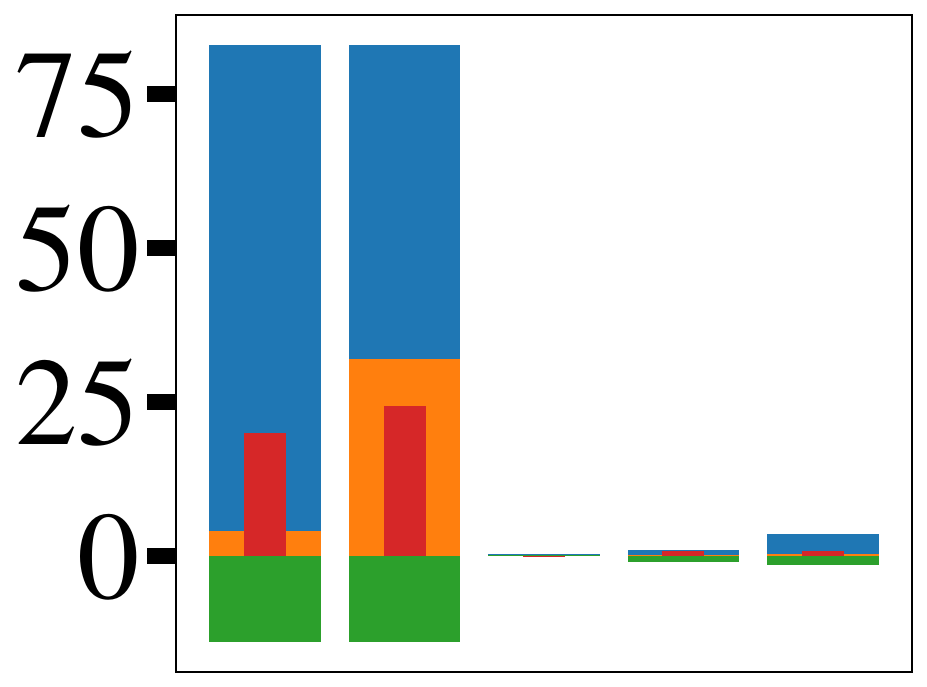}
        }{\centering Image not found}
        \vspace{-0.5cm}
        \caption{}\label{fig:input_parameter_details_b}
    \end{subfigure}
    \hfill
    \begin{subfigure}[b]{0.22\linewidth}
        \IfFileExists{figures/sic_no_wyckoff_interstitial_input_parameter_details_unscaled_single.pdf}{
            \includegraphics[width=\linewidth]{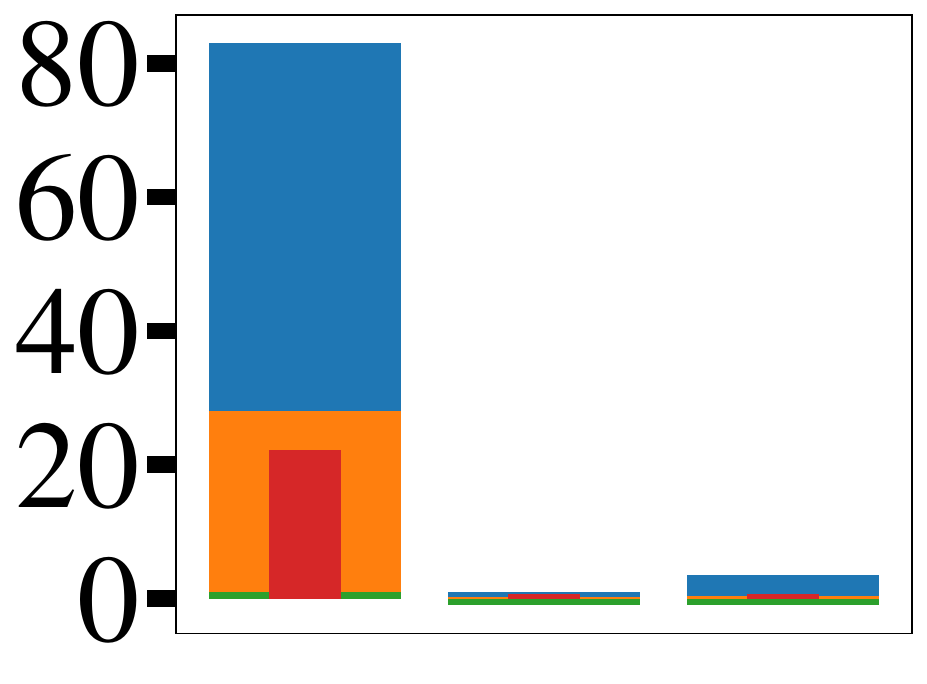}
        }{\centering Image not found}
        \vspace{-0.5cm}
        \caption{}\label{fig:input_parameter_details_c}
    \end{subfigure}
    \hfill
    \begin{subfigure}[b]{0.22\linewidth}
        \IfFileExists{figures/sic_no_wyckoff_interstitial_input_parameter_details_unscaled_double.pdf}{
            \includegraphics[width=\linewidth]{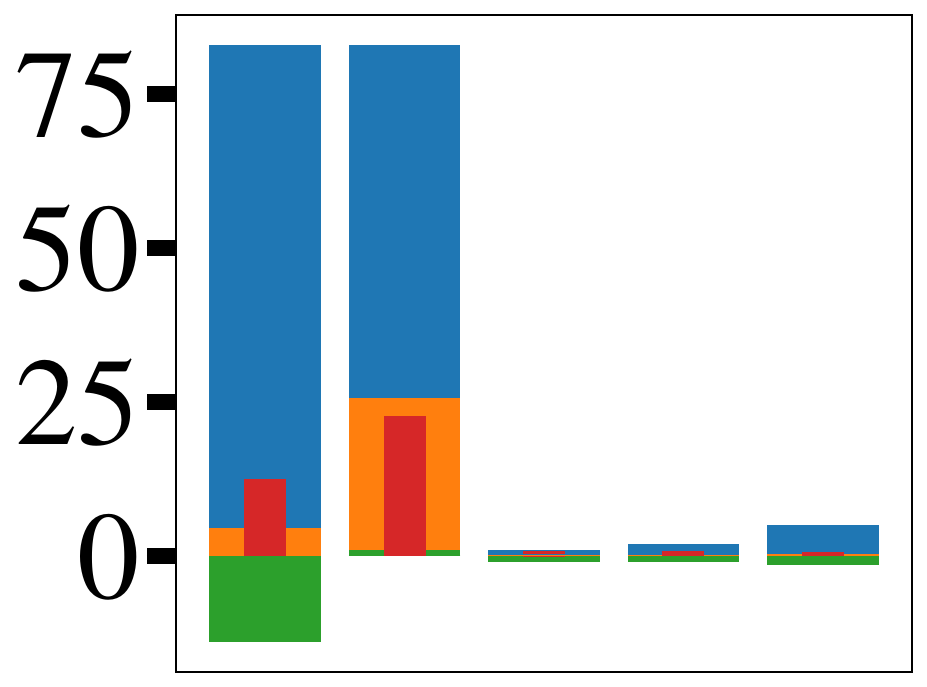}
        }{\centering Image not found}
        \vspace{-0.5cm}
        \caption{}\label{fig:input_parameter_details_d}
    \end{subfigure}


    \begin{subfigure}[b]{0.22\linewidth}
        \IfFileExists{figures/sic_wyckoff_letter_vacancy-substitutional_input_parameter_details_unscaled_single.pdf}{
            \includegraphics[width=\linewidth]{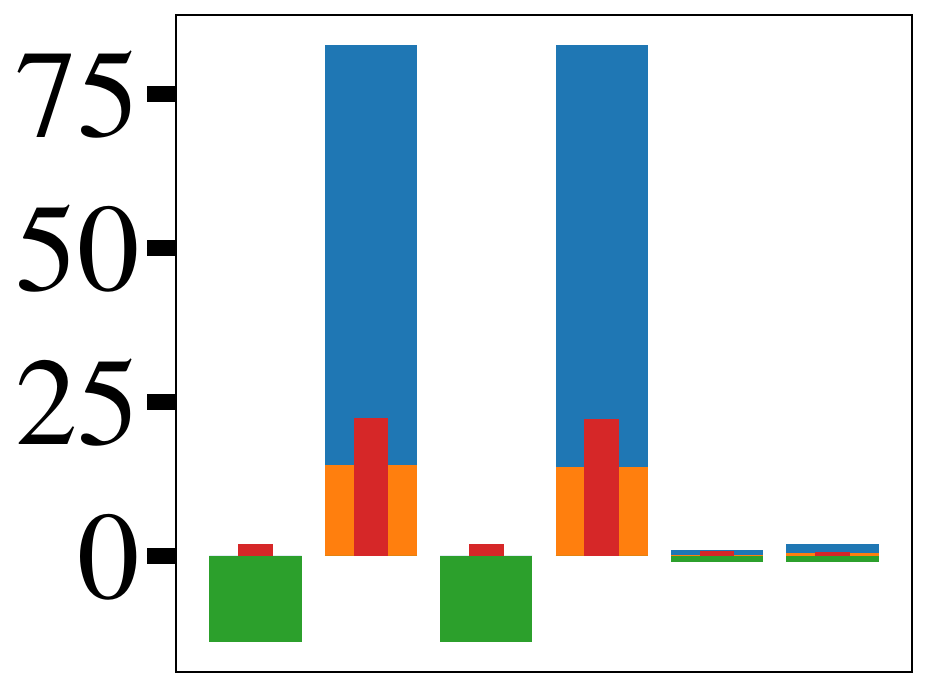}
        }{\centering Image not found}
        \vspace{-0.5cm}
        \caption{}\label{fig:input_parameter_details_e}
    \end{subfigure}
    \hfill
    \begin{subfigure}[b]{0.22\linewidth}
        \IfFileExists{figures/sic_wyckoff_letter_vacancy-substitutional_input_parameter_details_unscaled_double.pdf}{
            \includegraphics[width=\linewidth]{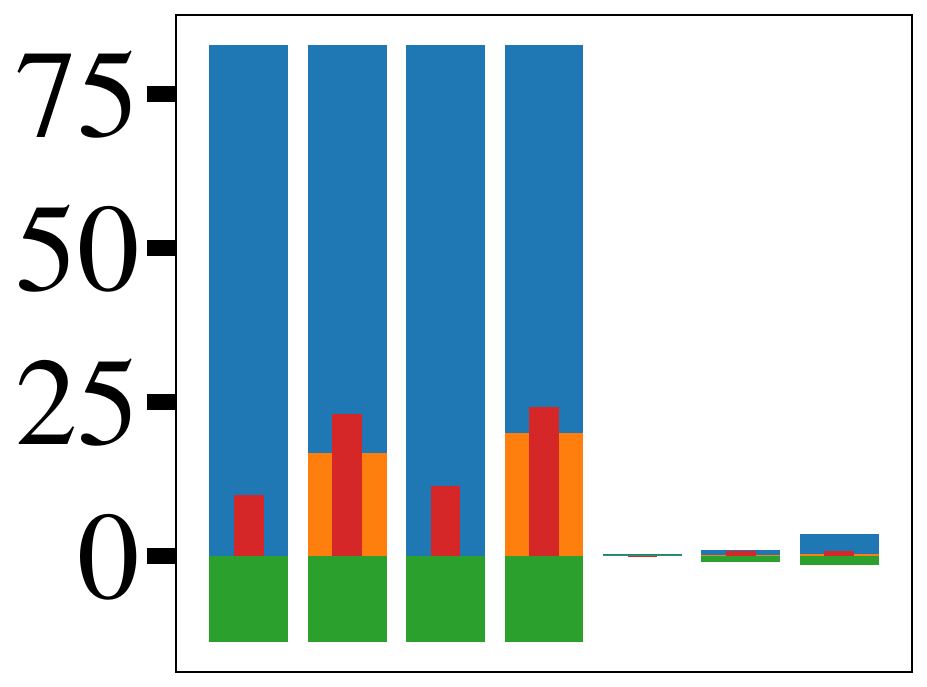}
        }{\centering Image not found}
        \vspace{-0.5cm}
        \caption{}\label{fig:input_parameter_details_f}
    \end{subfigure}
    \hfill
    \begin{subfigure}[b]{0.22\linewidth}
        \IfFileExists{figures/sic_wyckoff_letter_interstitial_input_parameter_details_unscaled_single.pdf}{
            \includegraphics[width=\linewidth]{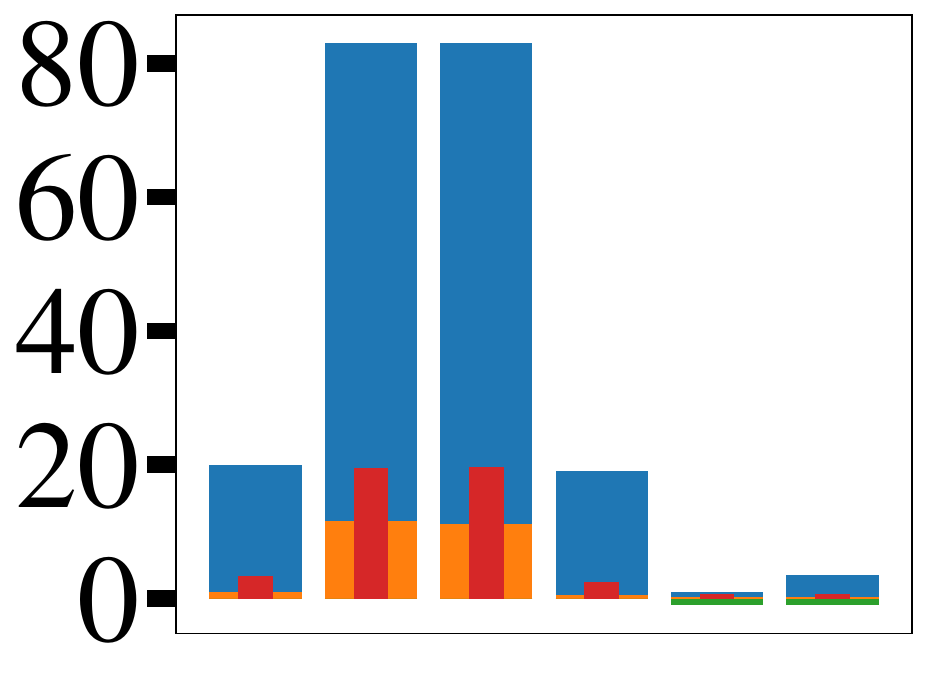}
        }{\centering Image not found}
        \vspace{-0.5cm}
        \caption{}\label{fig:input_parameter_details_g}
    \end{subfigure}
    \hfill
    \begin{subfigure}[b]{0.22\linewidth}
        \IfFileExists{figures/sic_wyckoff_letter_interstitial_input_parameter_details_unscaled_double.pdf}{
            \includegraphics[width=\linewidth]{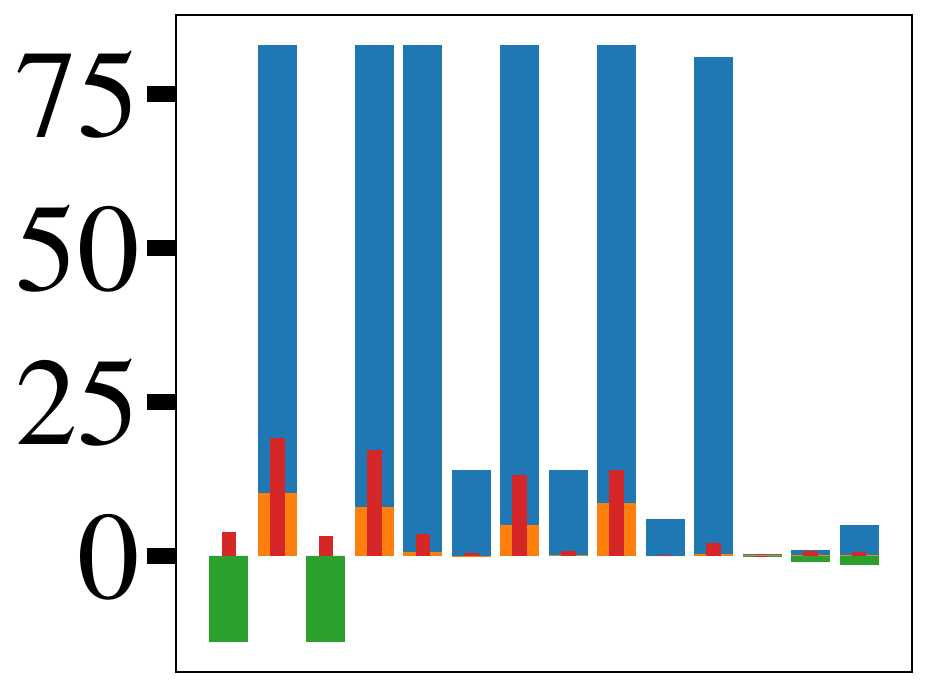}
        }{\centering Image not found}
        \vspace{-0.5cm}
        \caption{}\label{fig:input_parameter_details_h}
    \end{subfigure}

    \begin{subfigure}[b]{0.22\linewidth}
        \IfFileExists{figures/sic_wyckoff_vacancy-substitutional_input_parameter_details_unscaled_single.pdf}{
            \includegraphics[width=\linewidth]{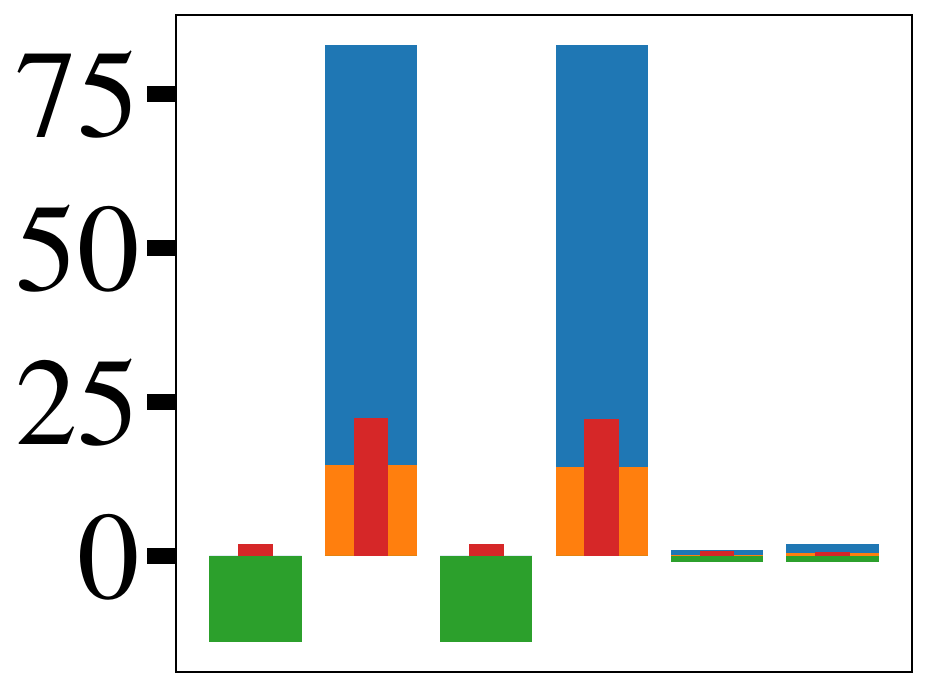}
        }{\centering Image not found}
        \vspace{-0.5cm}
        \caption{}\label{fig:input_parameter_details_i}
    \end{subfigure}
    \hfill
    \begin{subfigure}[b]{0.22\linewidth}
        \IfFileExists{figures/sic_wyckoff_vacancy-substitutional_input_parameter_details_unscaled_double.pdf}{
            \includegraphics[width=\linewidth]{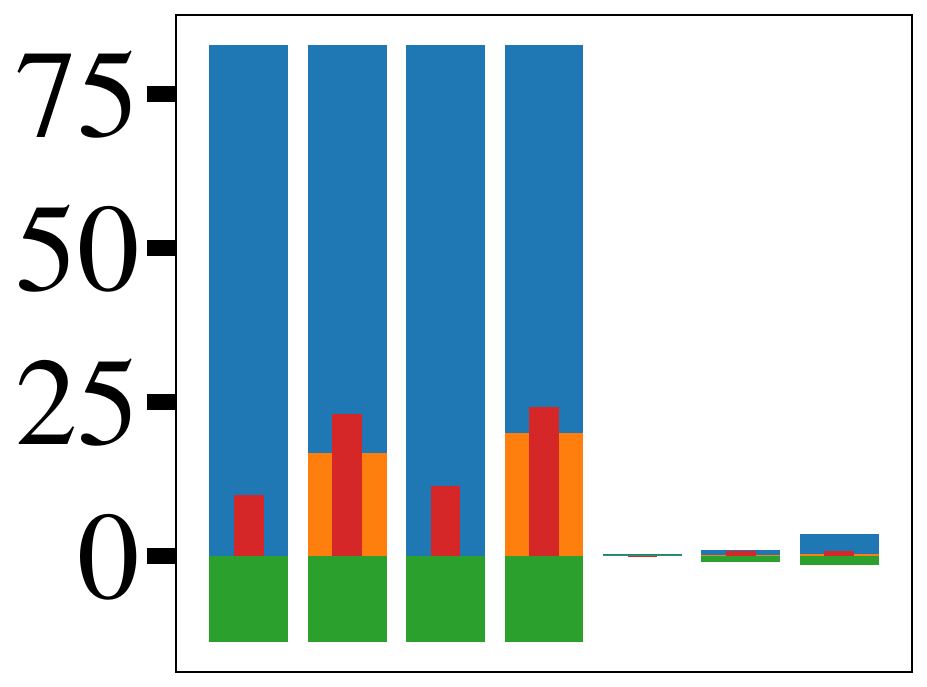}
        }{\centering Image not found}
        \vspace{-0.5cm}
        \caption{}\label{fig:input_parameter_details_j}
    \end{subfigure}
    \hfill
    \begin{subfigure}[b]{0.22\linewidth}
        \IfFileExists{figures/sic_wyckoff_interstitial_input_parameter_details_unscaled_single.pdf}{
            \includegraphics[width=\linewidth]{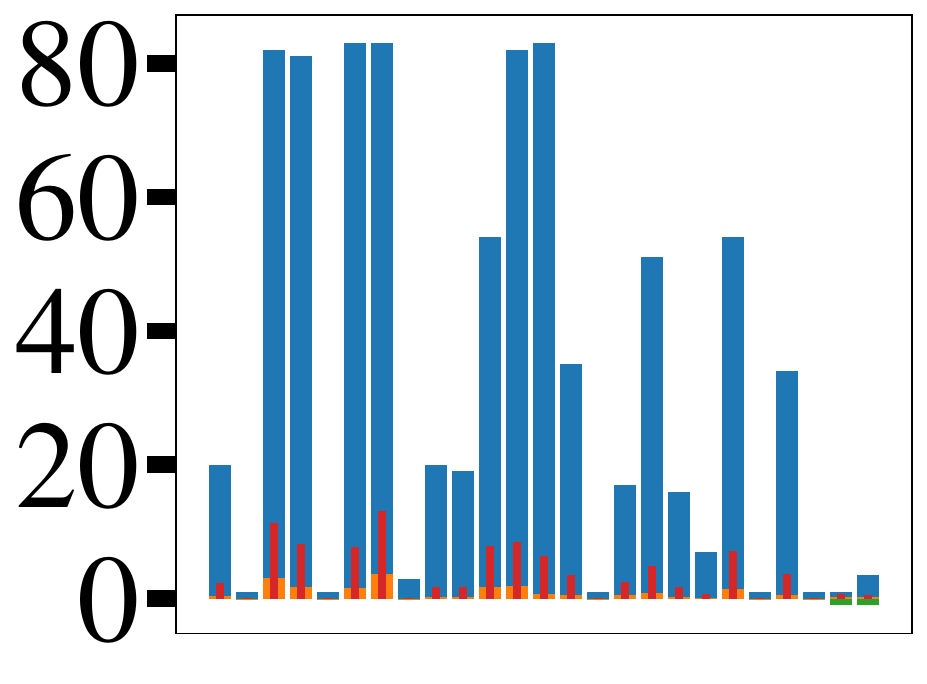}
        }{\centering Image not found}
        \vspace{-0.5cm}
        \caption{}\label{fig:input_parameter_details_k}
    \end{subfigure}
    \hfill
    \begin{subfigure}[b]{0.22\linewidth}
        \IfFileExists{figures/sic_wyckoff_interstitial_input_parameter_details_unscaled_double.pdf}{
            \includegraphics[width=\linewidth]{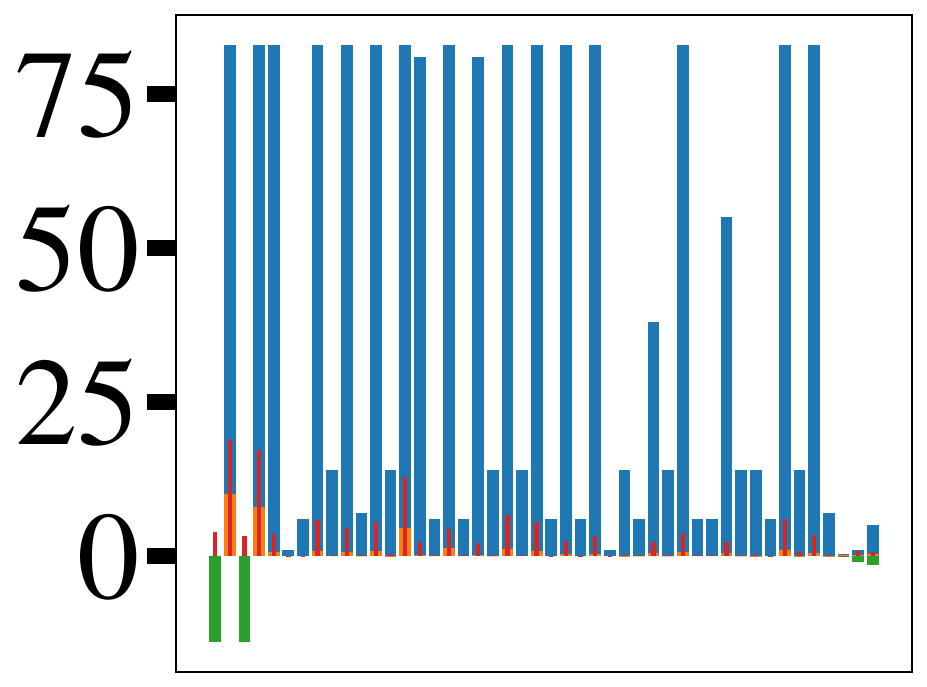}
        }{\centering Image not found}
        \vspace{-0.5cm}
        \caption{}\label{fig:input_parameter_details_l}
    \end{subfigure}

    \begin{subfigure}[b]{0.22\linewidth}
        \IfFileExists{figures/sic_wyckoff_letter_nw_data_vacancy-substitutional_input_parameter_details_unscaled_single.pdf}{
            \includegraphics[width=\linewidth]{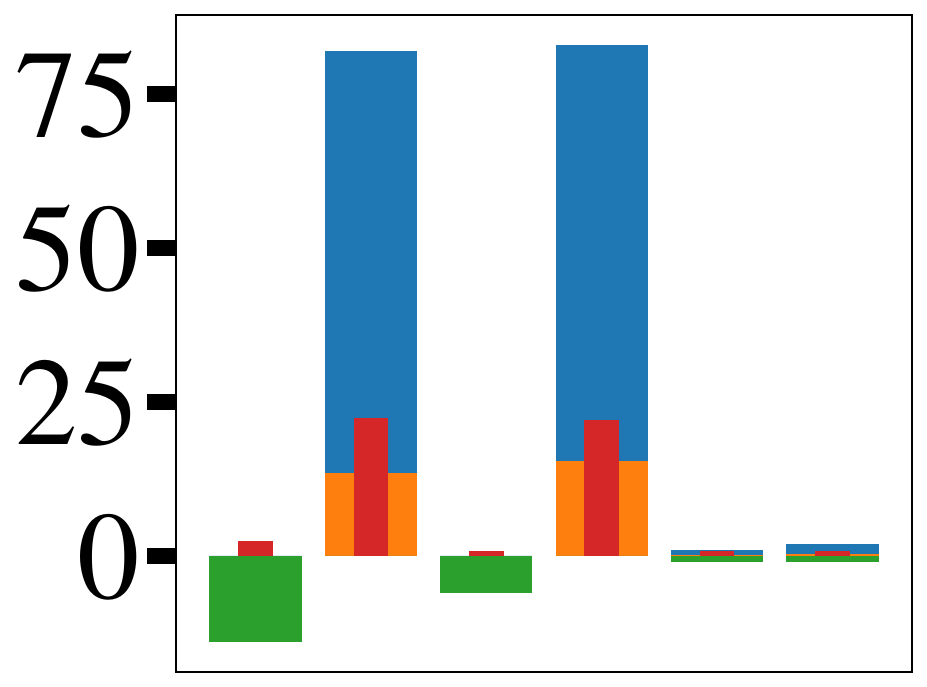}
        }{\centering Image not found}
        \vspace{-0.5cm}
        \caption{}\label{fig:input_parameter_details_m}
    \end{subfigure}
    \hfill
    \begin{subfigure}[b]{0.22\linewidth}
        \IfFileExists{figures/sic_wyckoff_letter_nw_data_vacancy-substitutional_input_parameter_details_unscaled_double.pdf}{
            \includegraphics[width=\linewidth]{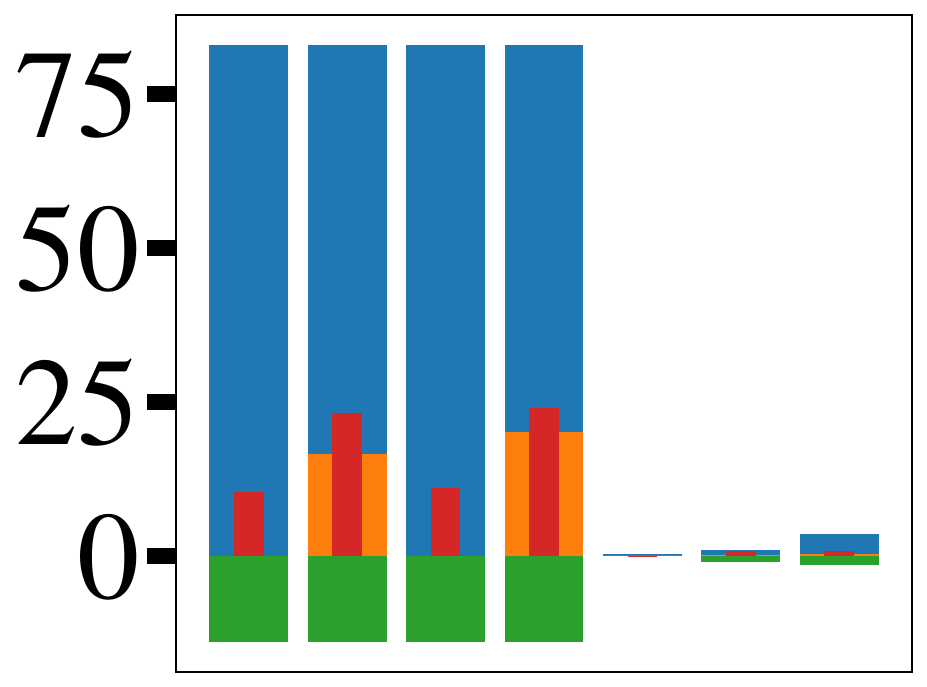}
        }{\centering Image not found}
        \vspace{-0.5cm}
        \caption{}\label{fig:input_parameter_details_n}
    \end{subfigure}
    \hfill
    \begin{subfigure}[b]{0.22\linewidth}
        \IfFileExists{figures/sic_wyckoff_letter_nw_data_interstitial_input_parameter_details_unscaled_single.pdf}{
            \includegraphics[width=\linewidth]{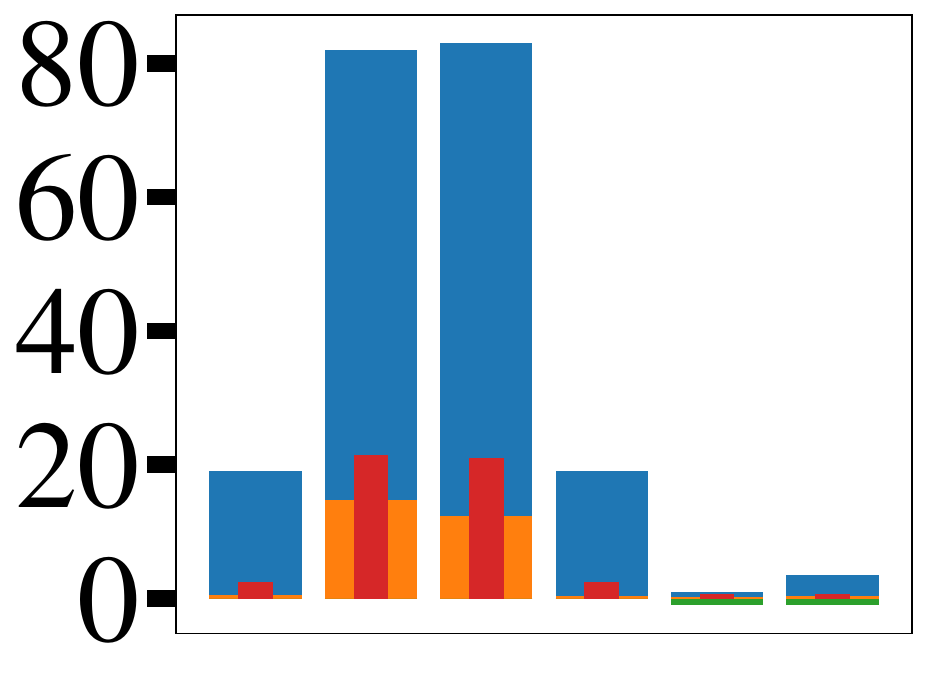}
        }{\centering Image not found}
        \vspace{-0.5cm}
        \caption{}\label{fig:input_parameter_details_o}
    \end{subfigure}
    \hfill
    \begin{subfigure}[b]{0.22\linewidth}
        \IfFileExists{figures/sic_wyckoff_letter_nw_data_interstitial_input_parameter_details_unscaled_double.pdf}{
            \includegraphics[width=\linewidth]{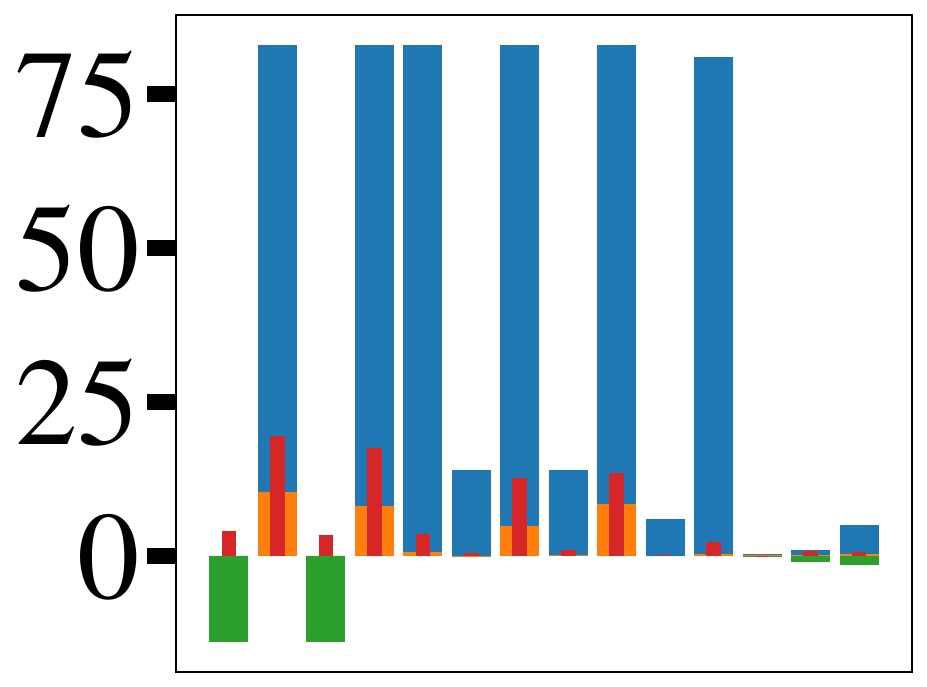}
        }{\centering Image not found}
        \vspace{-0.5cm}
        \caption{}\label{fig:input_parameter_details_p}
    \end{subfigure}

    \begin{subfigure}[b]{0.22\linewidth}
        \IfFileExists{figures/sic_wyckoff_nw_data_vacancy-substitutional_input_parameter_details_unscaled_single.pdf}{
            \includegraphics[width=\linewidth]{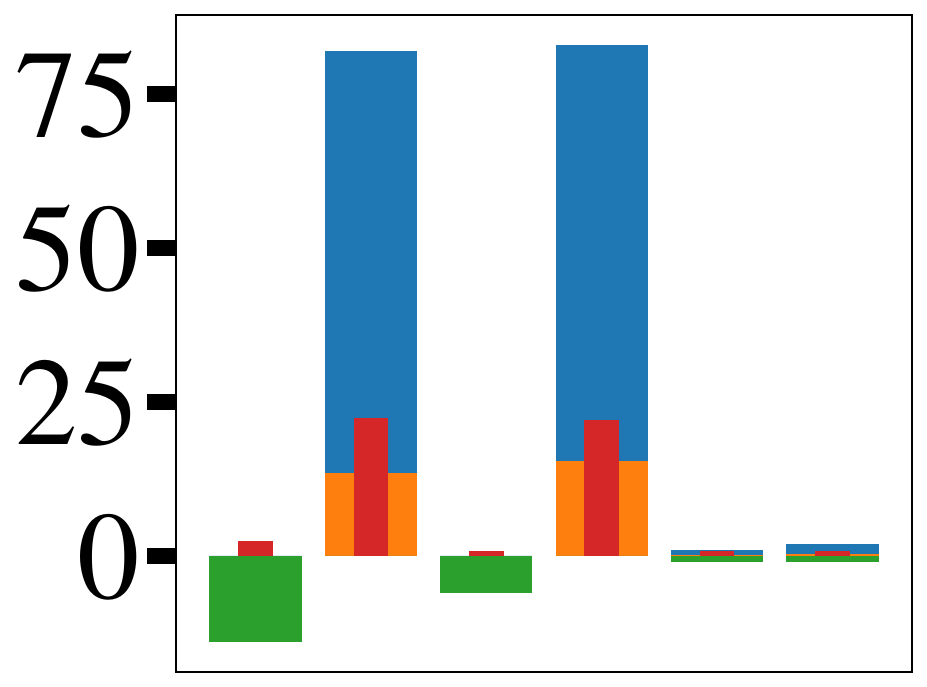}
        }{\centering Image not found}
        \vspace{-0.5cm}
        \caption{}\label{fig:input_parameter_details_q}
    \end{subfigure}
    \hfill
    \begin{subfigure}[b]{0.22\linewidth}
        \IfFileExists{figures/sic_wyckoff_nw_data_vacancy-substitutional_input_parameter_details_unscaled_double.pdf}{
            \includegraphics[width=\linewidth]{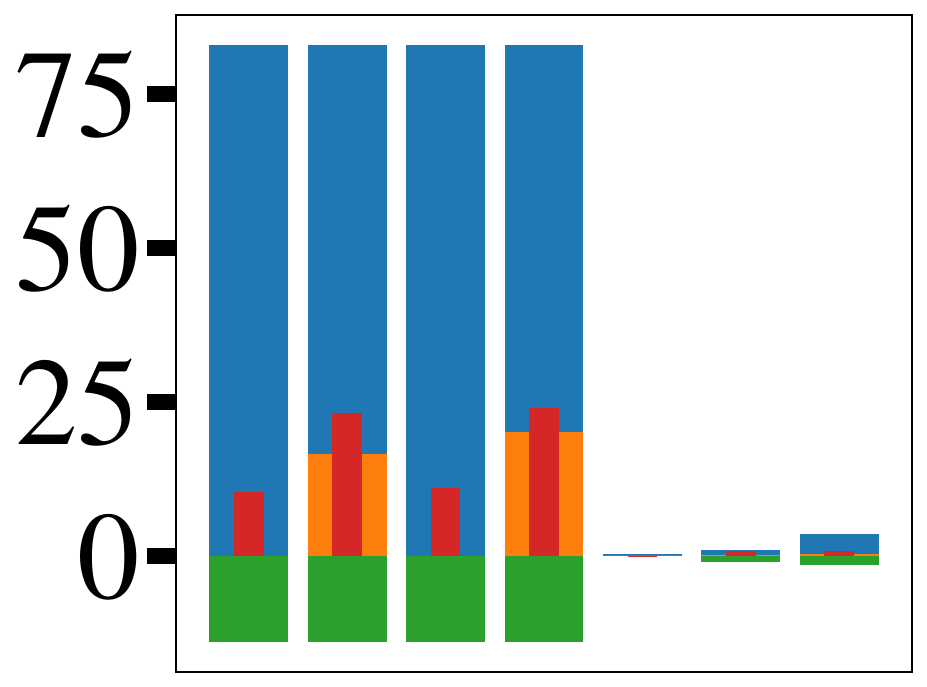}
        }{\centering Image not found}
        \vspace{-0.5cm}
        \caption{}\label{fig:input_parameter_details_r}
    \end{subfigure}
    \hfill
    \begin{subfigure}[b]{0.22\linewidth}
        \IfFileExists{figures/sic_wyckoff_nw_data_interstitial_input_parameter_details_unscaled_single.pdf}{
            \includegraphics[width=\linewidth]{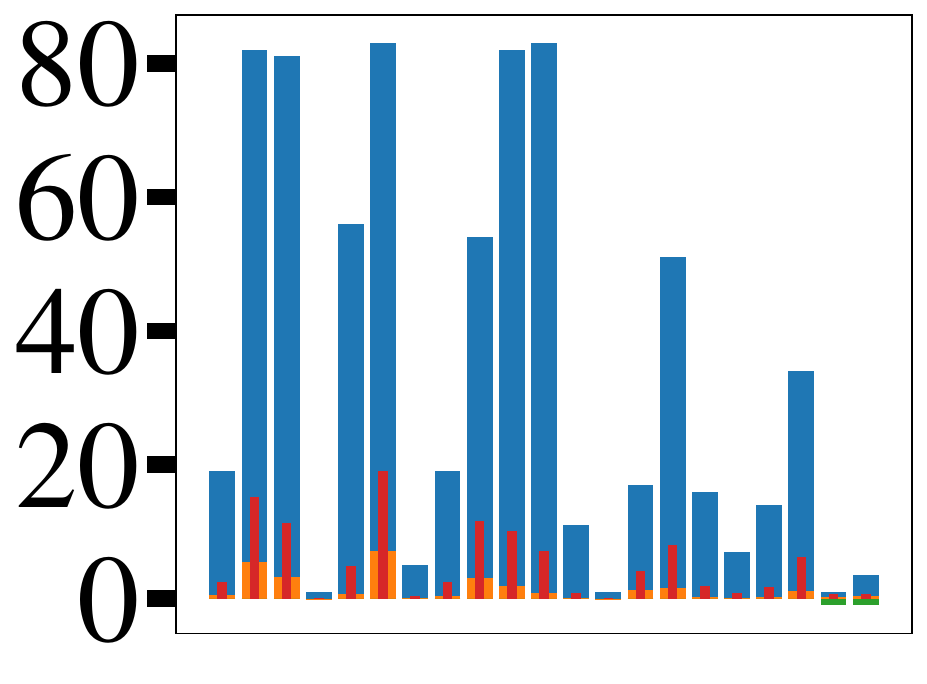}
        }{\centering Image not found}
        \vspace{-0.5cm}
        \caption{}\label{fig:input_parameter_details_s}
    \end{subfigure}
    \hfill
    \begin{subfigure}[b]{0.22\linewidth}
        \IfFileExists{figures/sic_wyckoff_nw_data_interstitial_input_parameter_details_unscaled_double.pdf}{
            \includegraphics[width=\linewidth]{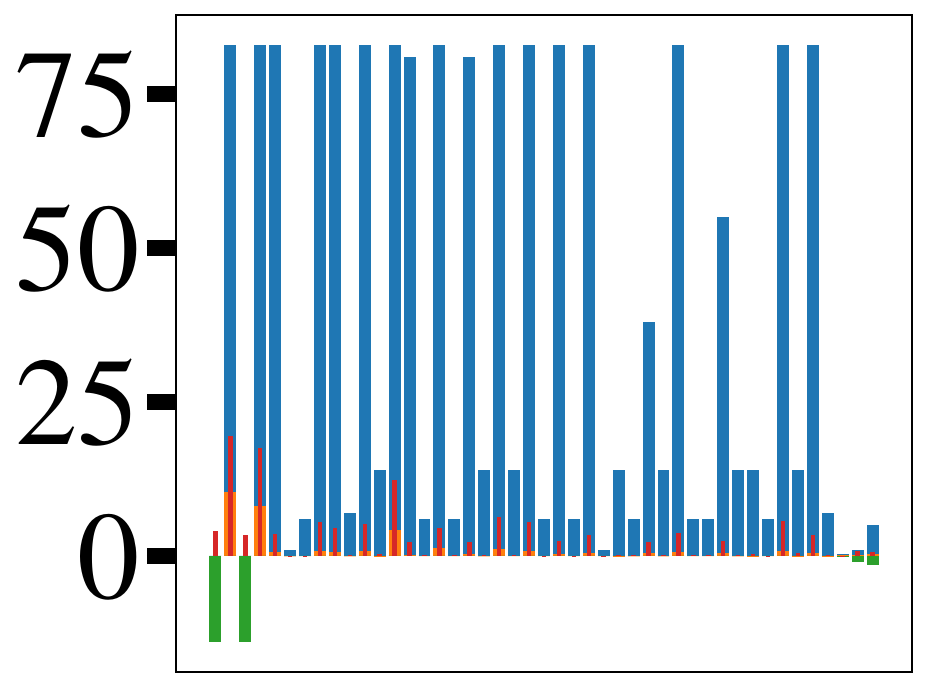}
        }{\centering Image not found}
        \vspace{-0.5cm}
        \caption{}\label{fig:input_parameter_details_t}
    \end{subfigure}

        \centering
        \normalsize Input Parameter
    \end{minipage}
}

    \caption{Input data set statistical properties for the datasets in Table~\ref{tab:dataset_details}, showing the minimum, maximum, average, and standard deviation. The datasets shown are: (a) A-VS-S, (b) A-VS-D, (c) A-I-S, (d) A-I-D, (e) WL-VS-S, (f) WL-VS-D, (g) WL-I-S, (h) WL-I-D, (i) WC-VS-S, (j) WC-VS-D, (k) WC-I-S, (l) WC-I-D, (m) WLC-VS-S, (n) WLC-VS-D, (o) WLC-I-S, (p) WLC-I-D, (q) WCC-VS-S, (r) WCC-VS-D, (s) WCC-I-S, (t) WCC-I-D.}
    \label{fig:input_parameter_details}
\end{figure}

The minimum, maximum, average, and standard deviation of each input parameter for each dataset are shown in Fig.~\ref{fig:input_parameter_details}. By looking at these values it is clear that the magnitude can vary significantly across input parameters. As stated in Section~\ref{sec:data_scaling}, ML models rarely perform well when the order of magnitude varies widely across input parameters. Therefore, it is motivated to scale the input parameters, especially in the case of the multilayer perceptron model since averages far from zero slow down the convergence, as discussed in Section~\ref{sec:data_scaling}. A slow convergence of the multilayer perceptron model is problematic, as it is already more sluggish and computationally intensive than the ridge and kernel ridge models. Additionally, standardizing the data has been shown to improve the accuracy of the multilayer perceptron model, as mentioned in Section~\ref{sec:data_scaling}.

\begin{figure}
    \centering


    \begin{subfigure}[b]{0.0924\linewidth}
        \IfFileExists{figures/sic_no_wyckoff_vacancy-substitutional_heatmap_covariance_unscaled.pdf}{
            \includegraphics[width=\linewidth]{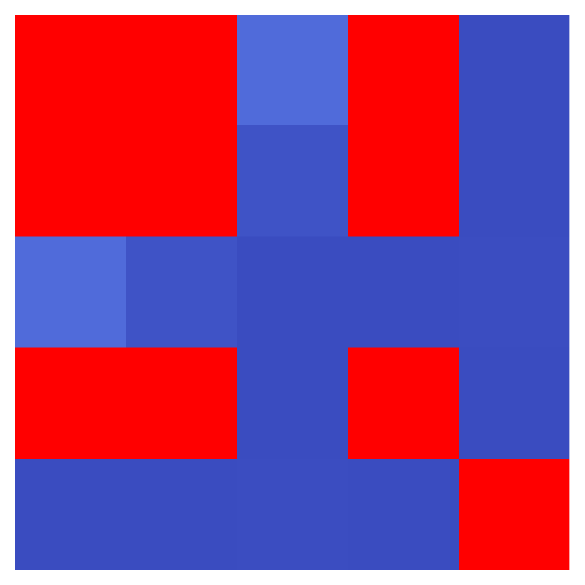}
        }{\centering Image not found}
        \vspace{-0.5cm}
        \caption{}\label{fig:heatmap_a}
    \end{subfigure}
    \hfill
    \begin{subfigure}[b]{0.0924\linewidth}
        \IfFileExists{figures/sic_no_wyckoff_interstitial_heatmap_covariance_unscaled.pdf}{
            \includegraphics[width=\linewidth]{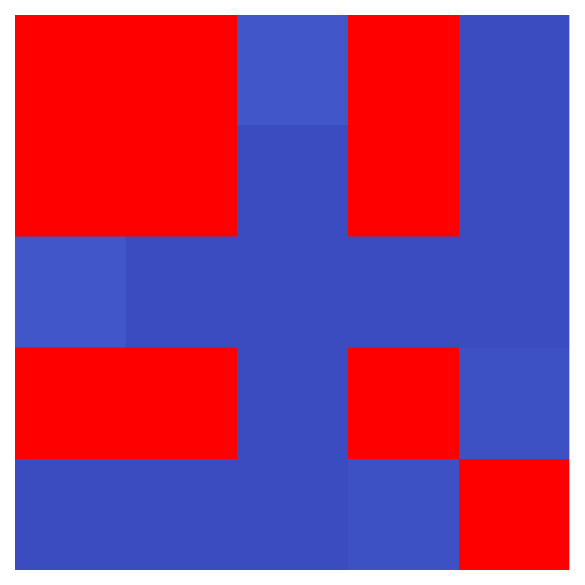}
        }{\centering Image not found}
        \vspace{-0.5cm}
        \caption{}\label{fig:heatmap_b}
    \end{subfigure}
    \hfill
    \begin{subfigure}[b]{0.0924\linewidth}
        \IfFileExists{figures/sic_wyckoff_letter_vacancy-substitutional_heatmap_covariance_unscaled.pdf}{
            \includegraphics[width=\linewidth]{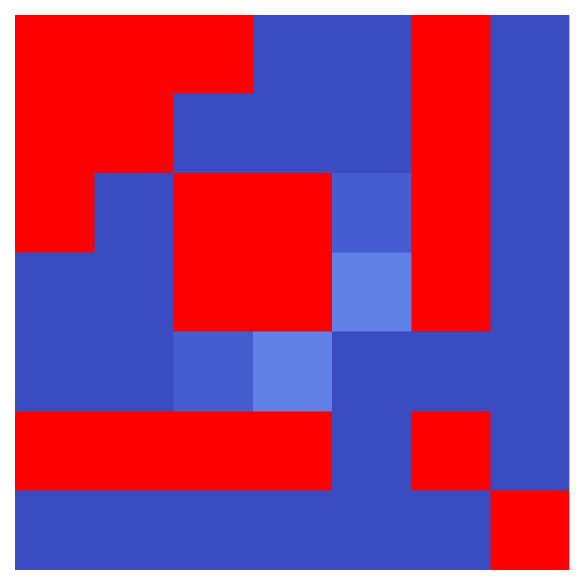}
        }{\centering Image not found}
        \vspace{-0.5cm}
        \caption{}\label{fig:heatmap_c}
    \end{subfigure}
    \hfill
    \begin{subfigure}[b]{0.0924\linewidth}
        \IfFileExists{figures/sic_wyckoff_letter_interstitial_heatmap_covariance_unscaled.pdf}{
            \includegraphics[width=\linewidth]{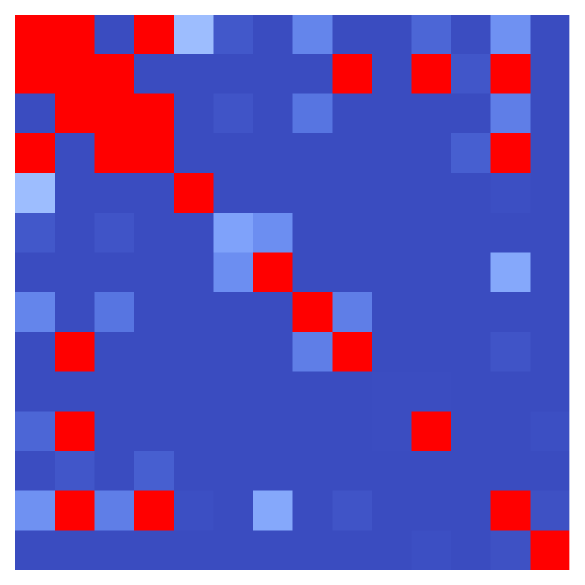}
        }{\centering Image not found}
        \vspace{-0.5cm}
        \caption{}\label{fig:heatmap_d}
    \end{subfigure}
    \hfill
    \begin{subfigure}[b]{0.0924\linewidth}
        \IfFileExists{figures/sic_wyckoff_vacancy-substitutional_heatmap_covariance_unscaled.pdf}{
            \includegraphics[width=\linewidth]{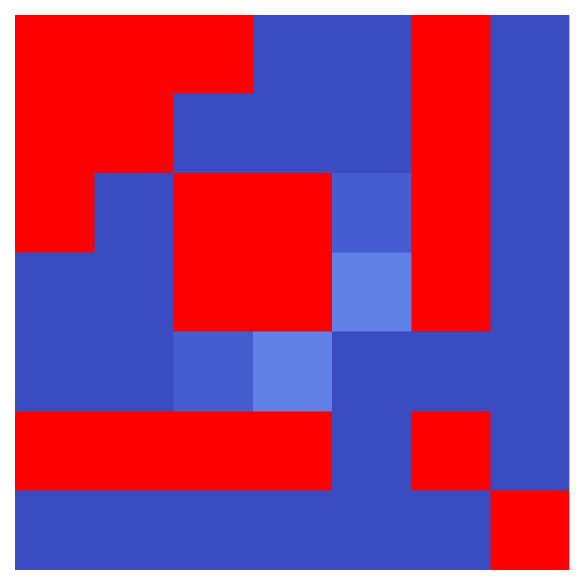}
        }{\centering Image not found}
        \vspace{-0.5cm}
        \caption{}\label{fig:heatmap_e}
    \end{subfigure}
    \hfill
    \begin{subfigure}[b]{0.0924\linewidth}
        \IfFileExists{figures/sic_wyckoff_interstitial_heatmap_covariance_unscaled.pdf}{
            \includegraphics[width=\linewidth]{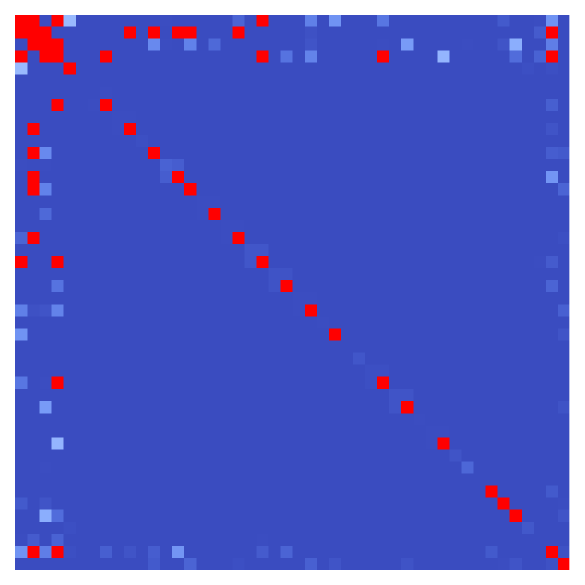}
        }{\centering Image not found}
        \vspace{-0.5cm}
        \caption{}\label{fig:heatmap_f}
    \end{subfigure}
    \hfill
    \begin{subfigure}[b]{0.0924\linewidth}
        \IfFileExists{figures/sic_wyckoff_letter_nw_data_vacancy-substitutional_heatmap_covariance_unscaled.pdf}{
            \includegraphics[width=\linewidth]{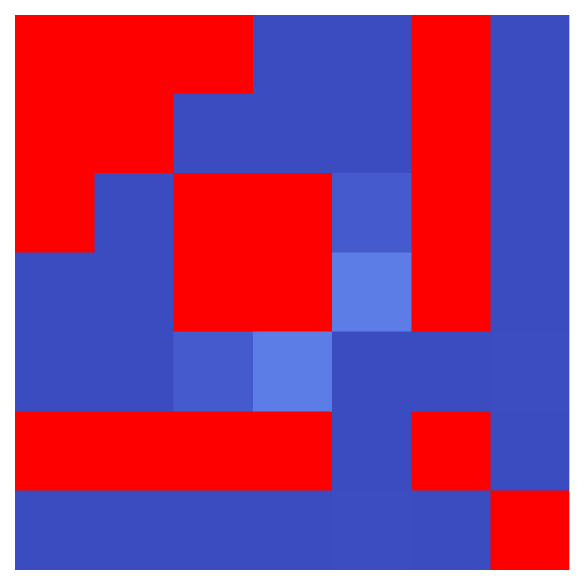}
        }{\centering Image not found}
        \vspace{-0.5cm}
        \caption{}\label{fig:heatmap_g}
    \end{subfigure}
    \hfill
    \begin{subfigure}[b]{0.0924\linewidth}
        \IfFileExists{figures/sic_wyckoff_letter_nw_data_interstitial_heatmap_covariance_unscaled.pdf}{
            \includegraphics[width=\linewidth]{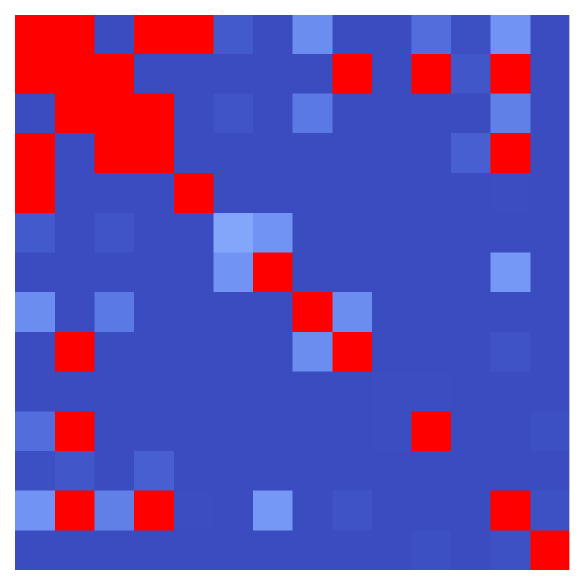}
        }{\centering Image not found}
        \vspace{-0.5cm}
        \caption{}\label{fig:heatmap_h}
    \end{subfigure}
    \hfill
    \begin{subfigure}[b]{0.0924\linewidth}
        \IfFileExists{figures/sic_wyckoff_nw_data_vacancy-substitutional_heatmap_covariance_unscaled.pdf}{
            \includegraphics[width=\linewidth]{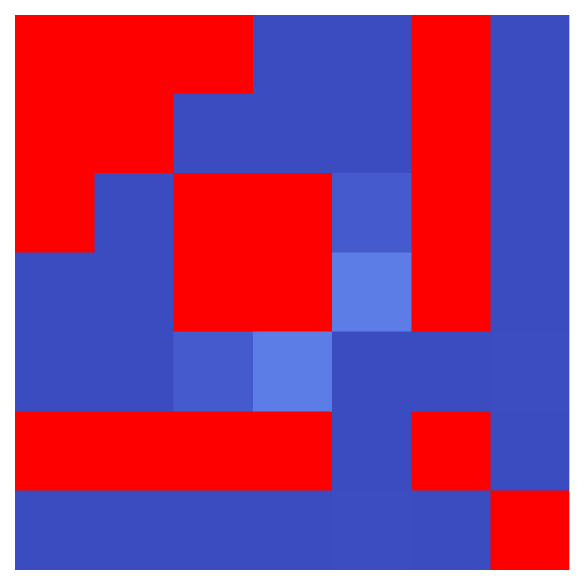}
        }{\centering Image not found}
        \vspace{-0.5cm}
        \caption{}\label{fig:heatmap_i}
    \end{subfigure}
    \hfill
    \begin{subfigure}[b]{0.0924\linewidth}
        \IfFileExists{figures/sic_wyckoff_nw_data_interstitial_heatmap_covariance_unscaled.pdf}{
            \includegraphics[width=\linewidth]{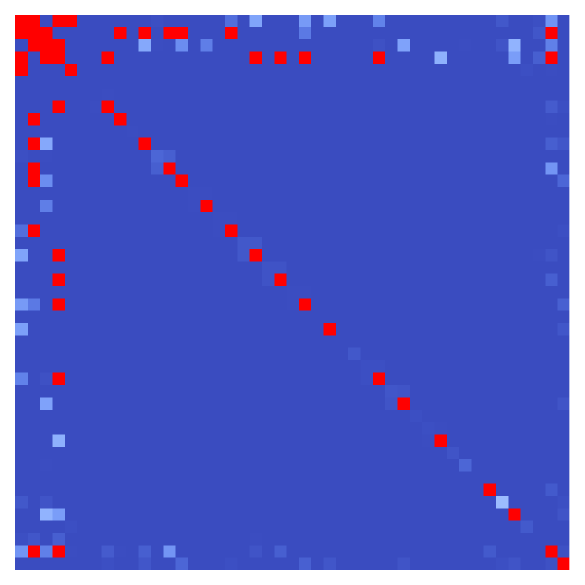}
        }{\centering Image not found}
        \vspace{-0.5cm}
        \caption{}\label{fig:heatmap_j}
    \end{subfigure}

    \begin{subfigure}[b]{0.0924\linewidth}
        \IfFileExists{figures/sic_no_wyckoff_vacancy-substitutional_heatmap_covariance_standardized.pdf}{
            \includegraphics[width=\linewidth]{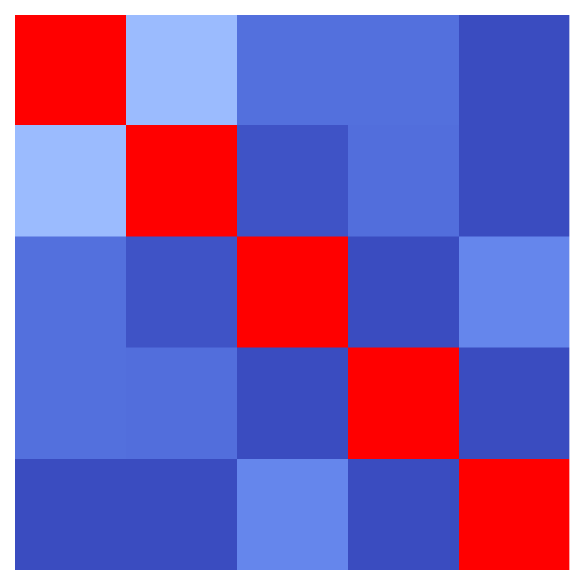}
        }{\centering Image not found}
        \vspace{-0.5cm}
        \caption{}\label{fig:heatmap_k}
    \end{subfigure}
    \hfill
    \begin{subfigure}[b]{0.0924\linewidth}
        \IfFileExists{figures/sic_no_wyckoff_interstitial_heatmap_covariance_standardized.pdf}{
            \includegraphics[width=\linewidth]{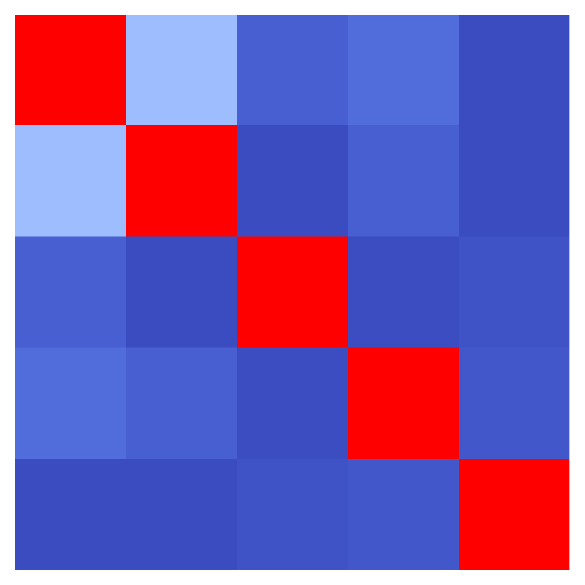}
        }{\centering Image not found}
        \vspace{-0.5cm}
        \caption{}\label{fig:heatmap_l}
    \end{subfigure}
    \hfill
    \begin{subfigure}[b]{0.0924\linewidth}
        \IfFileExists{figures/sic_wyckoff_letter_vacancy-substitutional_heatmap_covariance_standardized.pdf}{
            \includegraphics[width=\linewidth]{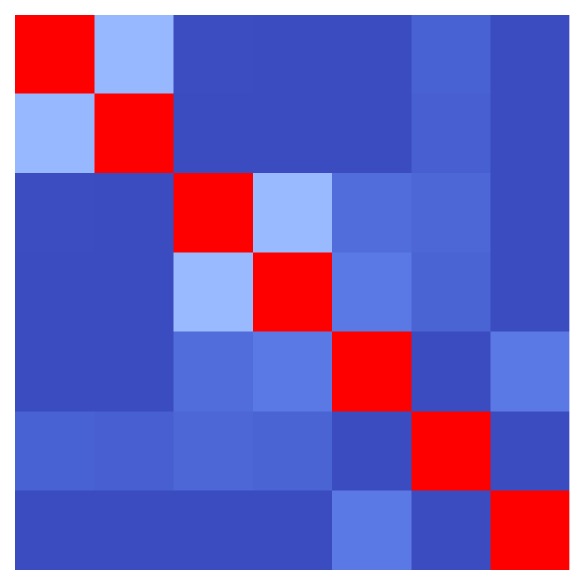}
        }{\centering Image not found}
        \vspace{-0.5cm}
        \caption{}\label{fig:heatmap_m}
    \end{subfigure}
    \hfill
    \begin{subfigure}[b]{0.0924\linewidth}
        \IfFileExists{figures/sic_wyckoff_letter_interstitial_heatmap_covariance_standardized.pdf}{
            \includegraphics[width=\linewidth]{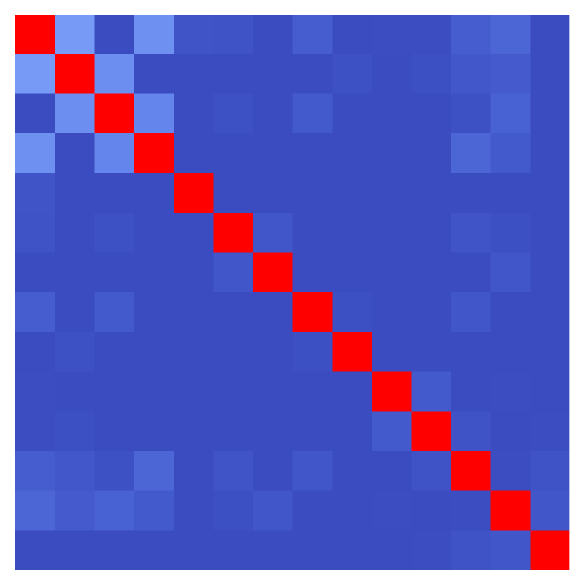}
        }{\centering Image not found}
        \vspace{-0.5cm}
        \caption{}\label{fig:heatmap_n}
    \end{subfigure}
    \hfill
    \begin{subfigure}[b]{0.0924\linewidth}
        \IfFileExists{figures/sic_wyckoff_vacancy-substitutional_heatmap_covariance_standardized.pdf}{
            \includegraphics[width=\linewidth]{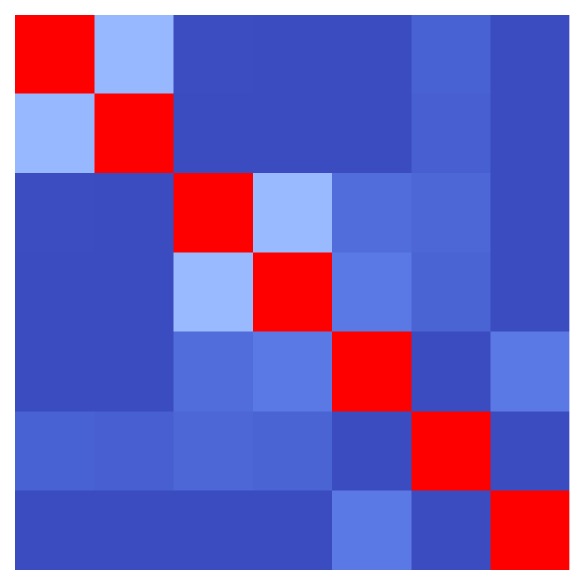}
        }{\centering Image not found}
        \vspace{-0.5cm}
        \caption{}\label{fig:heatmap_o}
    \end{subfigure}
    \hfill
    \begin{subfigure}[b]{0.0924\linewidth}
        \IfFileExists{figures/sic_wyckoff_interstitial_heatmap_covariance_standardized.pdf}{
            \includegraphics[width=\linewidth]{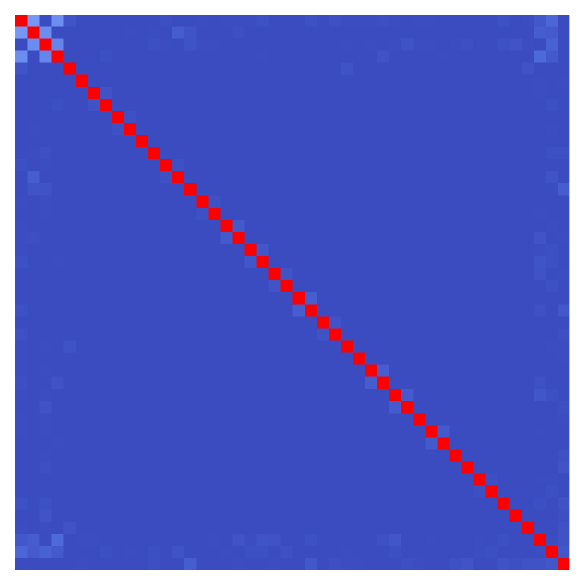}
        }{\centering Image not found}
        \vspace{-0.5cm}
        \caption{}\label{fig:heatmap_p}
    \end{subfigure}
    \hfill
    \begin{subfigure}[b]{0.0924\linewidth}
        \IfFileExists{figures/sic_wyckoff_letter_nw_data_vacancy-substitutional_heatmap_covariance_standardized.pdf}{
            \includegraphics[width=\linewidth]{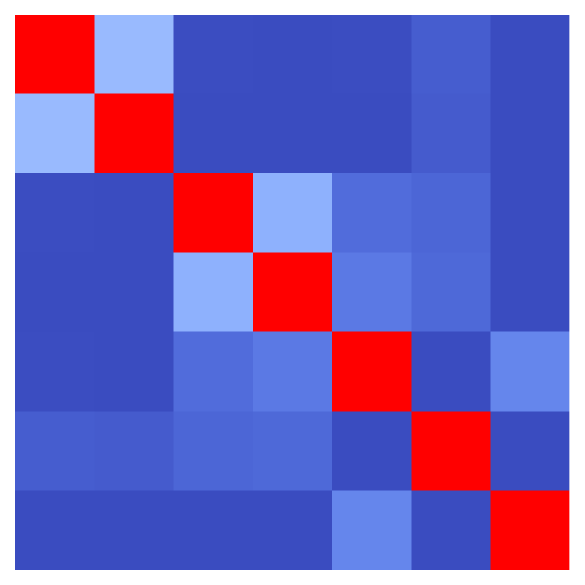}
        }{\centering Image not found}
        \vspace{-0.5cm}
        \caption{}\label{fig:heatmap_q}
    \end{subfigure}
    \hfill
    \begin{subfigure}[b]{0.0924\linewidth}
        \IfFileExists{figures/sic_wyckoff_letter_nw_data_interstitial_heatmap_covariance_standardized.pdf}{
            \includegraphics[width=\linewidth]{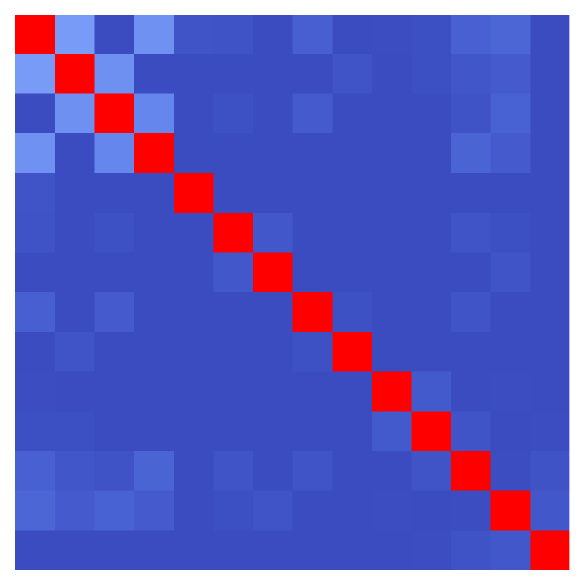}
        }{\centering Image not found}
        \vspace{-0.5cm}
        \caption{}\label{fig:heatmap_r}
    \end{subfigure}
    \hfill
    \begin{subfigure}[b]{0.0924\linewidth}
        \IfFileExists{figures/sic_wyckoff_nw_data_vacancy-substitutional_heatmap_covariance_standardized.pdf}{
            \includegraphics[width=\linewidth]{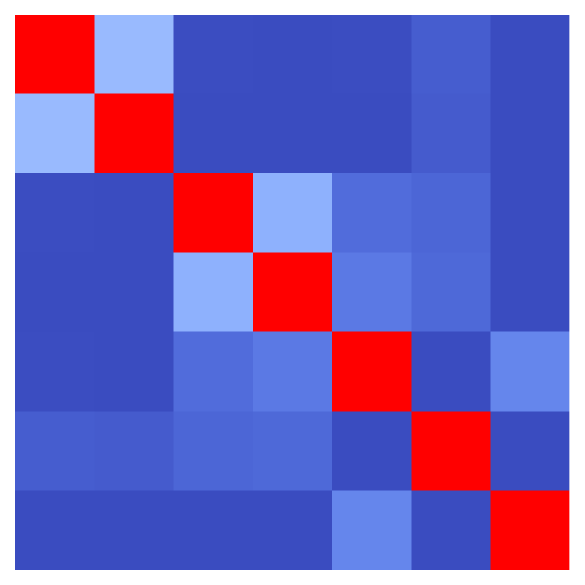}
        }{\centering Image not found}
        \vspace{-0.5cm}
        \caption{}\label{fig:heatmap_s}
    \end{subfigure}
    \hfill
    \begin{subfigure}[b]{0.0924\linewidth}
        \IfFileExists{figures/sic_wyckoff_nw_data_interstitial_heatmap_covariance_standardized.pdf}{
            \includegraphics[width=\linewidth]{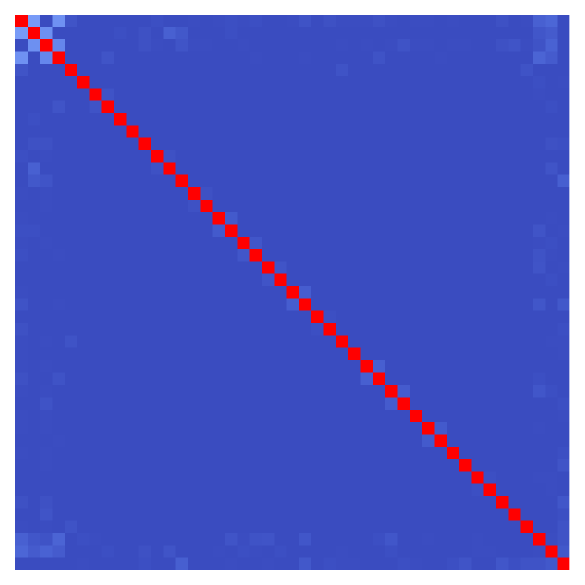}
        }{\centering Image not found}
        \vspace{-0.5cm}
        \caption{}\label{fig:heatmap_t}
    \end{subfigure}

    \begin{subfigure}[b]{0.0924\linewidth}
        \IfFileExists{figures/sic_no_wyckoff_vacancy-substitutional_heatmap_covariance_normalized.pdf}{
            \includegraphics[width=\linewidth]{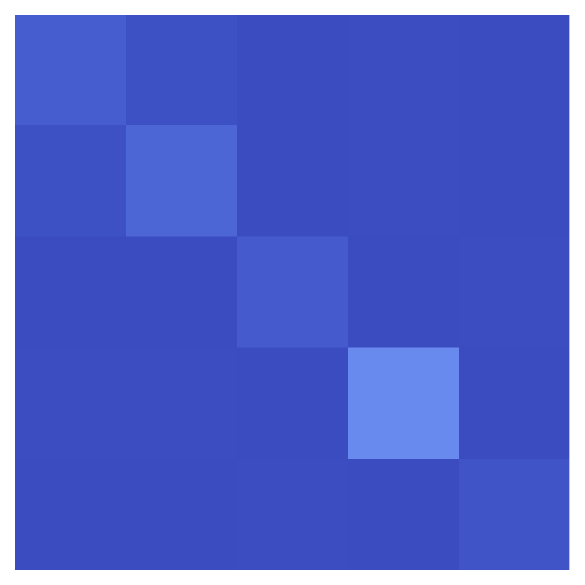}
        }{\centering Image not found}
        \vspace{-0.5cm}
        \caption{}\label{fig:heatmap_u}
    \end{subfigure}
    \hfill
    \begin{subfigure}[b]{0.0924\linewidth}
        \IfFileExists{figures/sic_no_wyckoff_interstitial_heatmap_covariance_normalized.pdf}{
            \includegraphics[width=\linewidth]{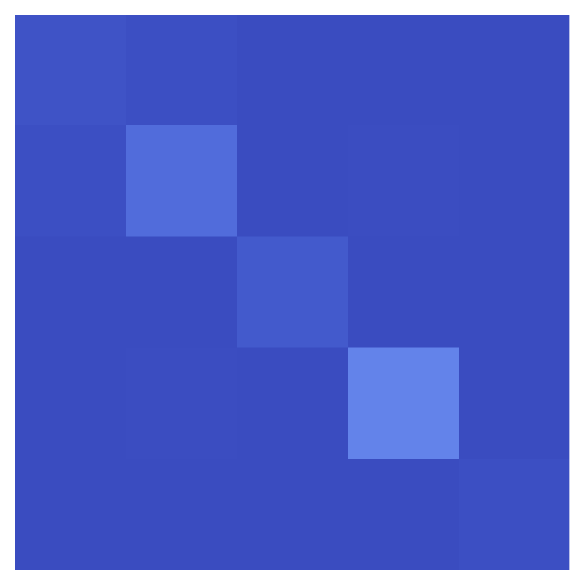}
        }{\centering Image not found}
        \vspace{-0.5cm}
        \caption{}\label{fig:heatmap_v}
    \end{subfigure}
    \hfill
    \begin{subfigure}[b]{0.0924\linewidth}
        \IfFileExists{figures/sic_wyckoff_letter_vacancy-substitutional_heatmap_covariance_normalized.pdf}{
            \includegraphics[width=\linewidth]{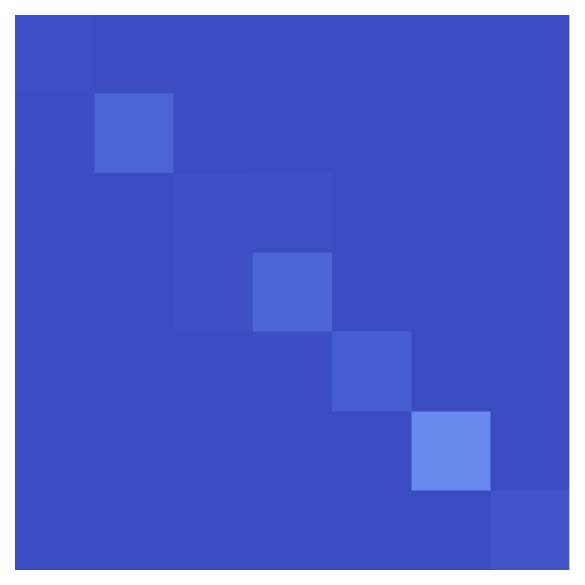}
        }{\centering Image not found}
        \vspace{-0.5cm}
        \caption{}\label{fig:heatmap_w}
    \end{subfigure}
    \hfill
    \begin{subfigure}[b]{0.0924\linewidth}
        \IfFileExists{figures/sic_wyckoff_letter_interstitial_heatmap_covariance_normalized.pdf}{
            \includegraphics[width=\linewidth]{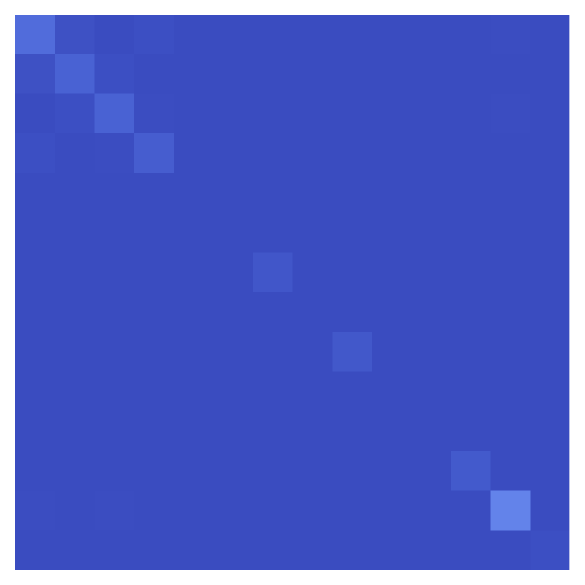}
        }{\centering Image not found}
        \vspace{-0.5cm}
        \caption{}\label{fig:heatmap_x}
    \end{subfigure}
    \hfill
    \begin{subfigure}[b]{0.0924\linewidth}
        \IfFileExists{figures/sic_wyckoff_vacancy-substitutional_heatmap_covariance_normalized.pdf}{
            \includegraphics[width=\linewidth]{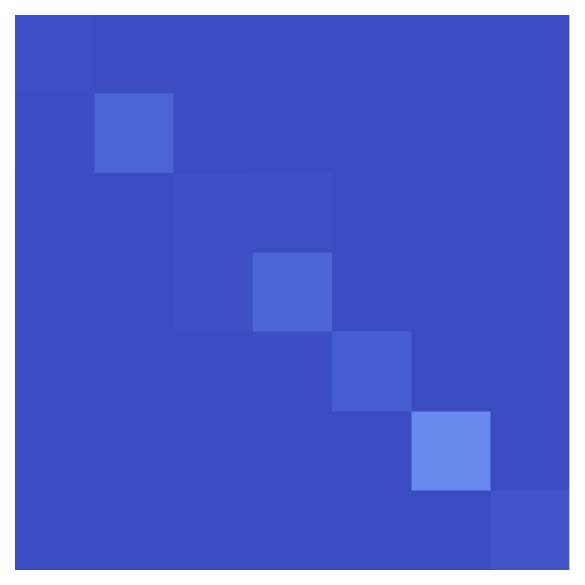}
        }{\centering Image not found}
        \vspace{-0.5cm}
        \caption{}\label{fig:heatmap_y}
    \end{subfigure}
    \hfill
    \begin{subfigure}[b]{0.0924\linewidth}
        \IfFileExists{figures/sic_wyckoff_interstitial_heatmap_covariance_normalized.pdf}{
            \includegraphics[width=\linewidth]{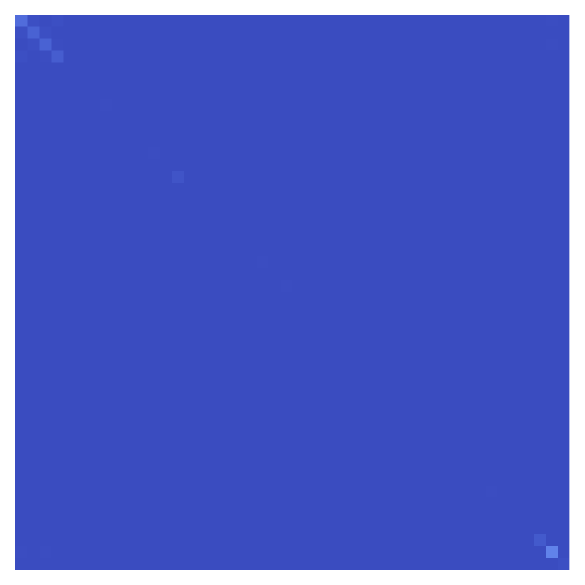}
        }{\centering Image not found}
        \vspace{-0.5cm}
        \caption{}\label{fig:heatmap_z}
    \end{subfigure}
    \hfill
    \begin{subfigure}[b]{0.0924\linewidth}
        \IfFileExists{figures/sic_wyckoff_letter_nw_data_vacancy-substitutional_heatmap_covariance_normalized.pdf}{
            \includegraphics[width=\linewidth]{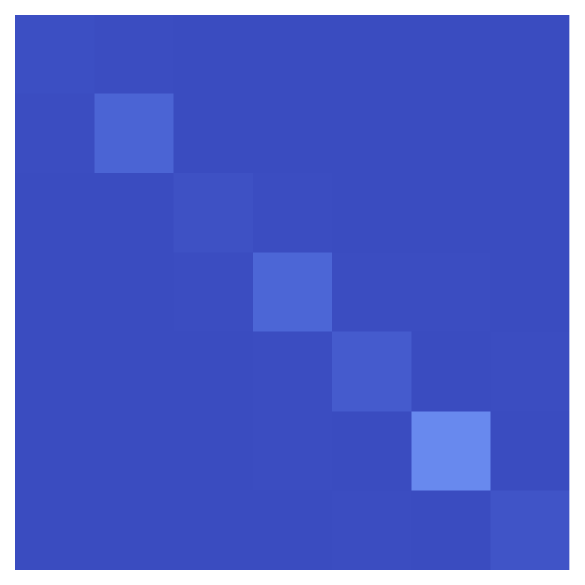}
        }{\centering Image not found}
        \vspace{-0.5cm}
        \caption{}\label{fig:heatmap_aa}
    \end{subfigure}
    \hfill
    \begin{subfigure}[b]{0.0924\linewidth}
        \IfFileExists{figures/sic_wyckoff_letter_nw_data_interstitial_heatmap_covariance_normalized.pdf}{
            \includegraphics[width=\linewidth]{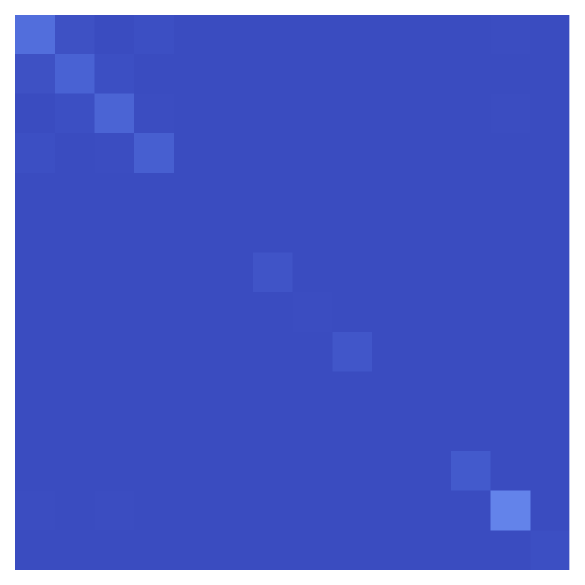}
        }{\centering Image not found}
        \vspace{-0.5cm}
        \caption{}\label{fig:heatmap_bb}
    \end{subfigure}
    \hfill
    \begin{subfigure}[b]{0.0924\linewidth}
        \IfFileExists{figures/sic_wyckoff_nw_data_vacancy-substitutional_heatmap_covariance_normalized.pdf}{
            \includegraphics[width=\linewidth]{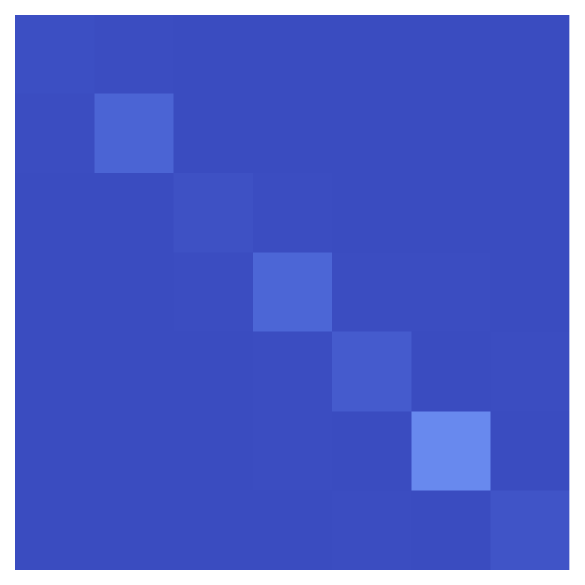}
        }{\centering Image not found}
        \vspace{-0.5cm}
        \caption{}\label{fig:heatmap_cc}
    \end{subfigure}
    \hfill
    \begin{subfigure}[b]{0.0924\linewidth}
        \IfFileExists{figures/sic_wyckoff_nw_data_interstitial_heatmap_covariance_normalized.pdf}{
            \includegraphics[width=\linewidth]{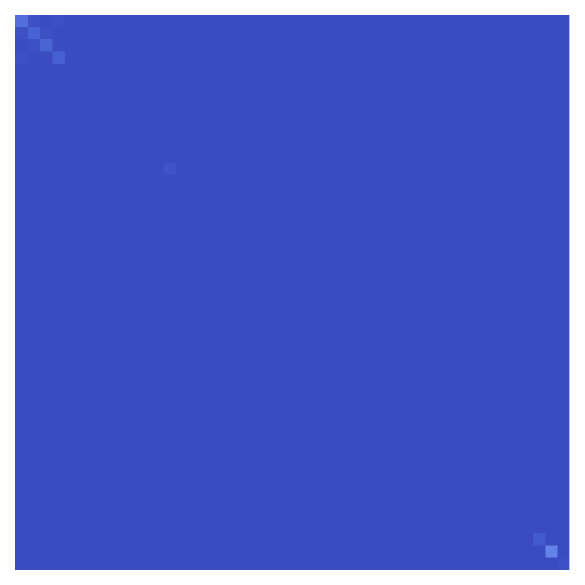}
        }{\centering Image not found}
        \vspace{-0.5cm}
        \caption{}\label{fig:heatmap_dd}
    \end{subfigure}

    \begin{subfigure}[b]{0.0924\linewidth}
        \IfFileExists{figures/sic_no_wyckoff_vacancy-substitutional_heatmap_pearson_coefficient.pdf}{
            \includegraphics[width=\linewidth]{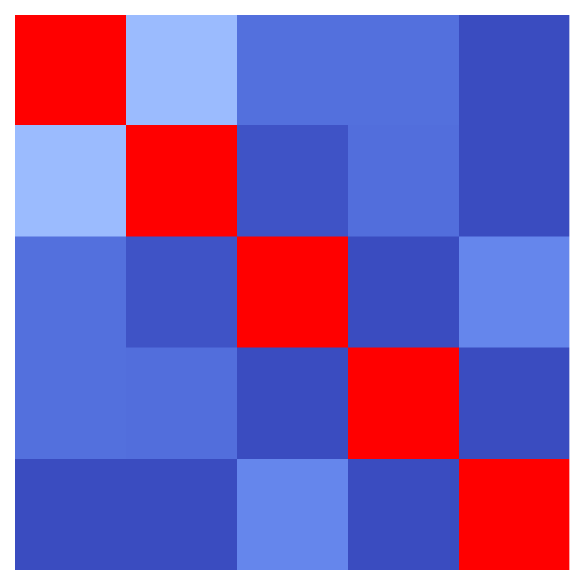}
        }{\centering Image not found}
        \vspace{-0.5cm}
        \caption{}\label{fig:heatmap_ee}
    \end{subfigure}
    \hfill
    \begin{subfigure}[b]{0.0924\linewidth}
        \IfFileExists{figures/sic_no_wyckoff_interstitial_heatmap_pearson_coefficient.pdf}{
            \includegraphics[width=\linewidth]{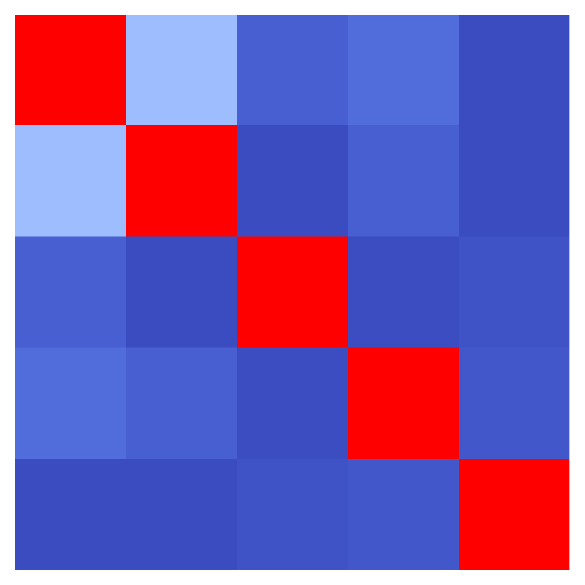}
        }{\centering Image not found}
        \vspace{-0.5cm}
        \caption{}\label{fig:heatmap_ff}
    \end{subfigure}
    \hfill
    \begin{subfigure}[b]{0.0924\linewidth}
        \IfFileExists{figures/sic_wyckoff_letter_vacancy-substitutional_heatmap_pearson_coefficient.pdf}{
            \includegraphics[width=\linewidth]{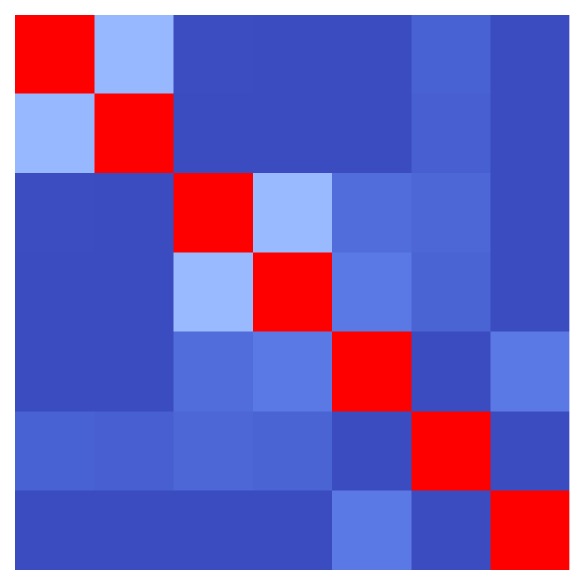}
        }{\centering Image not found}
        \vspace{-0.5cm}
        \caption{}\label{fig:heatmap_gg}
    \end{subfigure}
    \hfill
    \begin{subfigure}[b]{0.0924\linewidth}
        \IfFileExists{figures/sic_wyckoff_letter_interstitial_heatmap_pearson_coefficient.pdf}{
            \includegraphics[width=\linewidth]{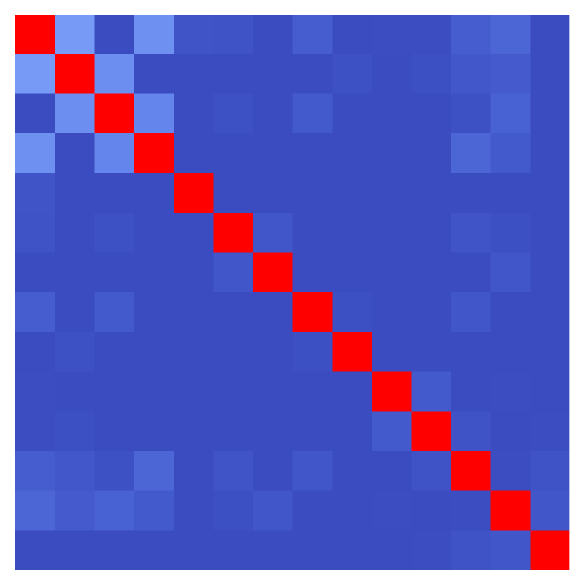}
        }{\centering Image not found}
        \vspace{-0.5cm}
        \caption{}\label{fig:heatmap_hh}
    \end{subfigure}
    \hfill
    \begin{subfigure}[b]{0.0924\linewidth}
        \IfFileExists{figures/sic_wyckoff_vacancy-substitutional_heatmap_pearson_coefficient.pdf}{
            \includegraphics[width=\linewidth]{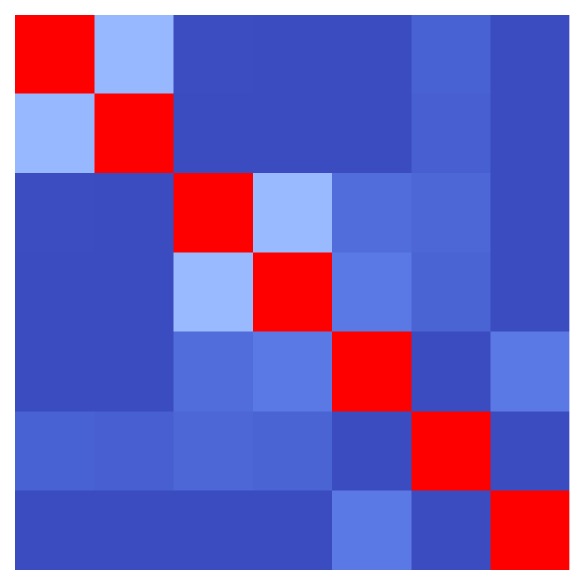}
        }{\centering Image not found}
        \vspace{-0.5cm}
        \caption{}\label{fig:heatmap_ii}
    \end{subfigure}
    \hfill
    \begin{subfigure}[b]{0.0924\linewidth}
        \IfFileExists{figures/sic_wyckoff_interstitial_heatmap_pearson_coefficient.pdf}{
            \includegraphics[width=\linewidth]{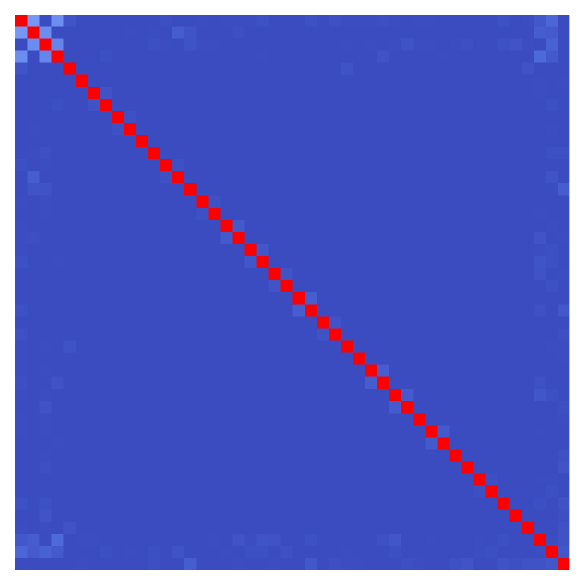}
        }{\centering Image not found}
        \vspace{-0.5cm}
        \caption{}\label{fig:heatmap_jj}
    \end{subfigure}
    \hfill
    \begin{subfigure}[b]{0.0924\linewidth}
        \IfFileExists{figures/sic_wyckoff_letter_nw_data_vacancy-substitutional_heatmap_pearson_coefficient.pdf}{
            \includegraphics[width=\linewidth]{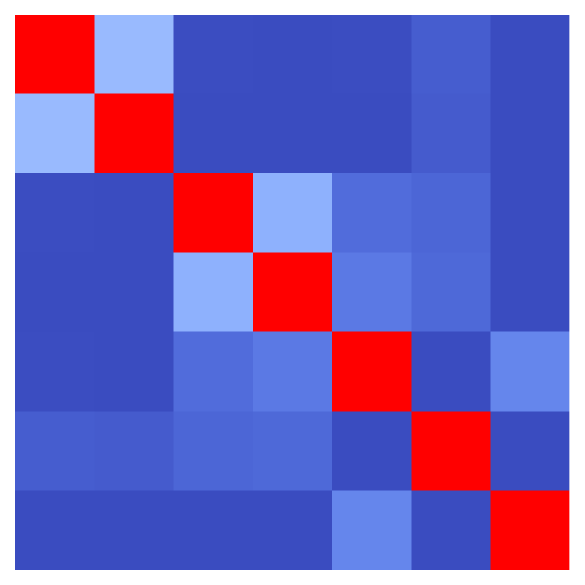}
        }{\centering Image not found}
        \vspace{-0.5cm}
        \caption{}\label{fig:heatmap_kk}
    \end{subfigure}
    \hfill
    \begin{subfigure}[b]{0.0924\linewidth}
        \IfFileExists{figures/sic_wyckoff_letter_nw_data_interstitial_heatmap_pearson_coefficient.pdf}{
            \includegraphics[width=\linewidth]{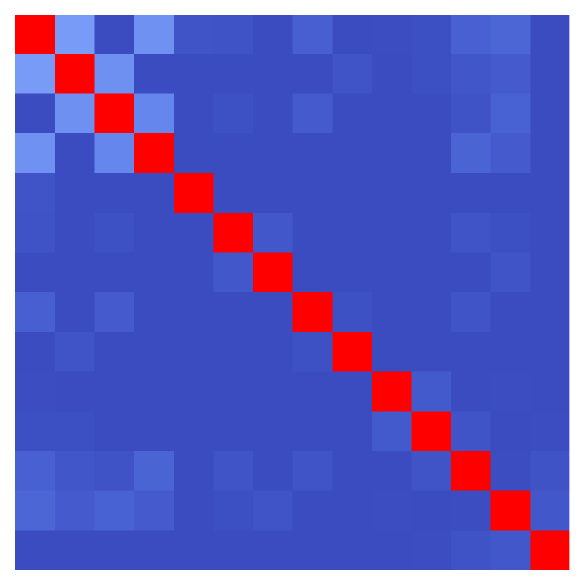}
        }{\centering Image not found}
        \vspace{-0.5cm}
        \caption{}\label{fig:heatmap_ll}
    \end{subfigure}
    \hfill
    \begin{subfigure}[b]{0.0924\linewidth}
        \IfFileExists{figures/sic_wyckoff_nw_data_vacancy-substitutional_heatmap_pearson_coefficient.pdf}{
            \includegraphics[width=\linewidth]{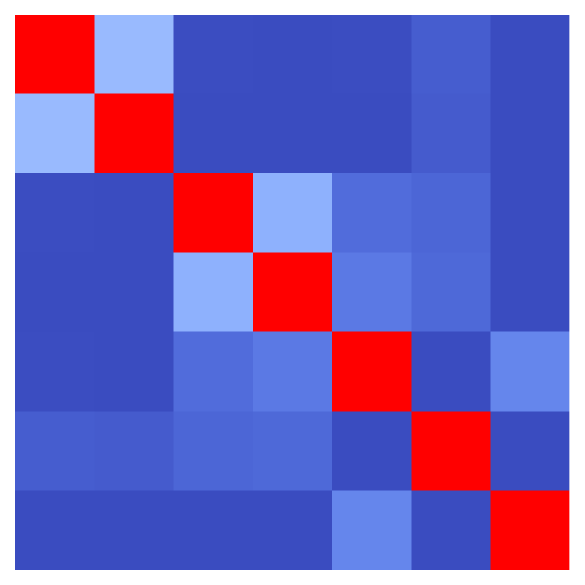}
        }{\centering Image not found}
        \vspace{-0.5cm}
        \caption{}\label{fig:heatmap_mm}
    \end{subfigure}
    \hfill
    \begin{subfigure}[b]{0.0924\linewidth}
        \IfFileExists{figures/sic_wyckoff_nw_data_interstitial_heatmap_pearson_coefficient.pdf}{
            \includegraphics[width=\linewidth]{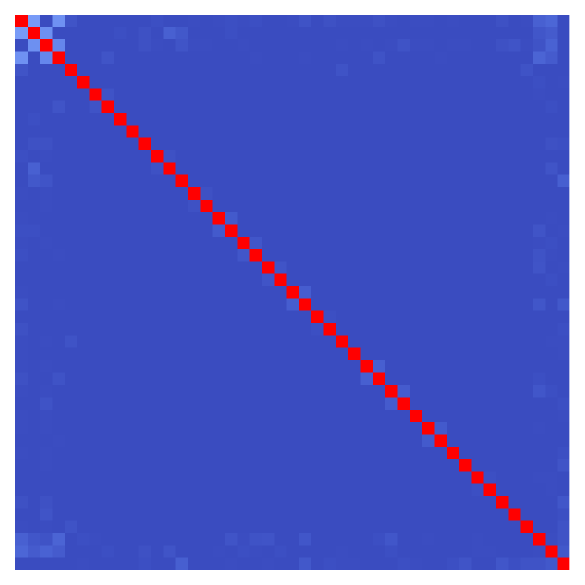}
        }{\centering Image not found}
        \vspace{-0.5cm}
        \caption{}\label{fig:heatmap_nn}
    \end{subfigure}


    \caption{Heatmaps of the correlation between input parameters in the datasets of Table~\ref{tab:dataset_details}. Magnitudes are shown in blue with darker shades for smaller magnitudes and values above \(0.31\) shown in red. Row 1 through 3 show: covariance between unscaled, standardized, and normalized input parameters. Row 4 shows the Pearson correlation coefficients between input parameters, which are identical across unscaled, standardized, and normalized input parameters. Columns 1 through 10 show the combined data of datasets: A-VS-S and A-VS-D, A-I-S and A-I-D, WL-VS-S and WL-VS-D, WL-I-S and WL-I-D, WC-VS-S and WC-VS-D, WC-I-S and WC-I-D, WLC-VS-S and WLC-VS-D, WLC-I-S and WLC-I-D, WCC-VS-S and WCC-VS-D, in addition to WCC-I-S and WCC-I-D.}
    \label{fig:heatmaps_correlation}
\end{figure}

Furthermore, the computed covariances and the Pearson correlation coefficients between all input parameters for each dataset are found in Fig.~\ref{fig:heatmaps_correlation}. Note that the input parameters are uncorrelated and that the covariances between input parameters are in general small. However, the covariances between some unscaled input parameters are in some cases moderately large, which slows down the convergence of the multilayer perceptron model, as discussed in Section~\ref{sec:data_scaling}. Scaling of the input parameters is therefore once more motivated.

\section{Results}\label{sec:results}

As described above, 10-fold cross-validation together with mean absolute error is used to calculate the final scores for each defect type, descriptor, scaling, and predicted energy. These final scores are given in Table~\ref{tab:all_scores} where the best ones for each defect type and predicted energy are highlighted. Note that, since in this work all single defects are included in training (see \ref{sec:performance_assessment}), these results only assess predictions of double defects.
For the combined vacancy–substitution category, the best MAEs are approximately 0.5 eV for formation energies and 0.2 eV for zero-phonon lines.
This result confidently allows for prescreening said point defects. The corresponding accuracies are around \(1.1\) and \(0.23\) eV for interstitials, indicating that ML can still be used for prescreening, but in a more coarse-grained manner. The corresponding ML hyperparameters that are used to obtain the results in Table~\ref{tab:all_scores} are given in Supplementary Materials Table S1.

\begin{table*}
\scriptsize
\begin{tabular}{ll
cccc|
cccccc|
cccc}
\toprule

& &
\multicolumn{4}{c|}{\textbf{Unscaled}} &
\multicolumn{6}{c|}{\textbf{Standardized}} &
\multicolumn{4}{c}{\textbf{Normalized}} \\

\cmidrule(r){3-6} \cmidrule(r){7-12} \cmidrule(r){13-16}

& &
\multicolumn{2}{c}{\textbf{Ridge}} &
\multicolumn{2}{c|}{\textbf{Kernel Ridge}} &
\multicolumn{2}{c}{\textbf{Ridge}} &
\multicolumn{2}{c}{\textbf{Kernel Ridge}} &
\multicolumn{2}{c|}{\textbf{Multilayer Perceptron}} &
\multicolumn{2}{c}{\textbf{Ridge}} &
\multicolumn{2}{c}{\textbf{Kernel Ridge}} \\

\cmidrule(r){3-4} \cmidrule(r){5-6}
\cmidrule(r){7-8} \cmidrule(r){9-10} \cmidrule(r){11-12}
\cmidrule(r){13-14} \cmidrule(r){15-16}

\textbf{Defect Type} & \textbf{Descriptor}
&
\textbf{FE} & \textbf{ZPL} &
\textbf{FE} & \textbf{ZPL} &
\textbf{FE} & \textbf{ZPL} &
\textbf{FE} & \textbf{ZPL} &
\textbf{FE} & \textbf{ZPL} &
\textbf{FE} & \textbf{ZPL} &
\textbf{FE} & \textbf{ZPL} \\

\midrule

\multirow{5}{*}{Vacancy-Substitution}
& Atomic
& 2.598 & 0.270 &  \textcolor{mplorange}{ \textbf{\underline{1.263}} } & \setlength{\fboxsep}{1.5pt}\fcolorbox{optimumcolor}{white}{ \textcolor{mplorange}{ \textbf{0.202} } }
& 2.598 & 0.270 & 2.003 & 0.210  & \setlength{\fboxsep}{1.5pt}\fcolorbox{optimumcolor}{white}{ \textcolor{mplgreen}{ \textbf{0.478} } } &  \textcolor{mplgreen}{ \textbf{\underline{0.234}} }
& 2.597 & 0.270 & 2.020 & 0.258  \\

& Wyckoff Letter
& 2.576 & 0.266 & 1.314 & 0.242
& 2.575 &  \textcolor{mplblue}{ \underline{\textbf{0.266}} } & 1.986 & 0.250  & 0.496 & 0.252
&  \textcolor{mplblue}{ \textbf{\underline{2.572}} } & 0.266 & 2.008 & 0.257  \\

& Wyckoff Coordinate
& 2.576 & 0.266 & 1.314 & 0.242
& 2.575 & 0.266 & 1.986 & 0.250  & 0.485 & 0.252
& 2.572 & 0.266 & 2.008 & 0.257  \\

& Wyckoff Letter Compare
& 2.595 & 0.270 & 1.314 & 0.248
& 2.594 & 0.270 & 2.004 & 0.259  & 0.528 & 0.262
& 2.592 & 0.270 & 2.021 & 0.262  \\

& Wyckoff Coordinate Compare
& 2.595 & 0.270 & 1.314 & 0.248
& 2.594 & 0.270 & 2.004 & 0.259  & 0.545 & 0.257
& 2.592 & 0.270 & 2.021 & 0.262  \\

\midrule

\multirow{5}{*}{Interstitial}
& Atomic
& 3.562 & 0.280 & 2.530 & \setlength{\fboxsep}{1.5pt}\fcolorbox{optimumcolor}{white}{ \textcolor{mplorange}{ \textbf{0.230} } }
& 3.559 & 0.280 & 3.072 & 0.254  & 2.582 &  \textcolor{mplgreen}{ \textbf{\underline{0.246}} }
& 3.554 & 0.280 & 3.047 & 0.259  \\

& Wyckoff Letter
& 3.462 & 0.281 & 2.420 & 0.232
& 3.460 & 0.281 & 2.851 & 0.259  & 1.523 & 0.247
& 3.461 & 0.281 & 2.880 & 0.265  \\

& Wyckoff Coordinate
& 3.301 & 0.280 &  \textcolor{mplorange}{ \textbf{\underline{1.818}} } & 0.237
&  \textcolor{mplblue}{ \underline{\textbf{3.301}} } & 0.280 & 2.214 & 0.260  & \setlength{\fboxsep}{1.5pt}\fcolorbox{optimumcolor}{white}{ \textcolor{mplgreen}{ \textbf{1.123} } } & 0.257
& 3.301 & 0.280 & 2.418 & 0.266  \\

& Wyckoff Letter Compare
& 3.482 & 0.279 & 2.410 & 0.230
& 3.480 & 0.279 & 2.850 & 0.257  & 1.633 & 0.248
& 3.481 & 0.279 & 2.882 & 0.263  \\

& Wyckoff Coordinate Compare
& 3.326 & 0.279 & 1.841 & 0.235
& 3.326 & 0.279 & 2.225 & 0.259  & 1.182 & 0.258
& 3.326 &  \textcolor{mplblue}{ \textbf{\underline{0.279}} } & 2.422 & 0.265  \\

\bottomrule
\end{tabular}
\centering
\caption{The best 10-fold cross-validation MAE for each predicted energy, defect type, descriptor, scaling, and machine learning model. The overall best MAEs are highlighted with red boxes, and for the others, the best for each model type is underlined: ridge model (blue), kernel ridge (orange), multilayer perceptron (green). More details on the corresponding hyperparameters are given in Supplementary Materials Table S1.}
\label{tab:all_results}
\label{tab:all_scores}
\end{table*}

Concerning the choice of descriptors, the most notable result is that the Atomic Descriptor achieves the best performance in three out of four cases, as highlighted in Table~\ref{tab:all_scores}. The exception is interstitial formation energy predictions, where the best performance is achieved using the Wyckoff Coordinate Descriptor.  The Wyckoff Letter Descriptor also improves interstitial formation-energy predictions when paired with the multilayer perceptron model, although its MAE remains noticeably higher than that obtained with the Wyckoff Coordinate Descriptor. Overall, the multilayer perceptron model with standardized input parameters achieves the best performance for formation-energy predictions, whereas the kernel ridge model with unscaled input parameters achieves the best performance for zero-phonon-line predictions.

Varied split ratio 10-fold cross-validation is used to plot the learning curves for each of the three types of ML models using their respective best combination of scaling, descriptor, and hyperparameter configuration given in Table~\ref{tab:all_scores} for each defect type and predicted energy. These learning curves for the test and training data are seen in Fig.~\ref{fig:mae_data_points_train}. These plots show that the test performance of each model generally improves as more training data points are provided, whereas the performance of the training data generally worsens. To clarify, this is also true for the ridge model, even though the performance variation for the test and training data is minimal. Moreover, Fig.~\ref{fig:mae_data_points_train_a}~and~\ref{fig:mae_data_points_train_c} show the performance variation of formation energy predictions, and it looks like the performance plateaus for a large enough number of data points. This plateauing behavior is rarely the case for other ML models across any of the defect types or predicted energies. The exception is for formation energy predictions in vacancies and substitutions, where the kernel ridge model also plateaus for a large enough number of training data points, which can be seen in Fig.~\ref{fig:mae_data_points_train_a}.

The projected scores calculated as the mean of the final train and test scores for each defect type and predicted energy, which correspond to Fig.~\ref{fig:mae_data_points_train}, are given in Table~\ref{tab:projections}. These projection scores further show that the multilayer perceptron model achieves the best performance for formation energy predictions and that the kernel ridge model achieves the best performance regarding zero-phonon line predictions.

\begin{table}
\centering
\begin{tabular}{ll|ccc}
\textbf{Defect Type} & \textbf{Energy} & \textbf{R} & \textbf{KR} & \textbf{MLP} \\
\midrule
\multirow{2}{*}{Vacancy-Subst.} & Formation Energy (eV) & 2.563 & 1.088 & 0.379 \\
& Zero-Phonon Line (eV) & 0.266 & 0.166 & 0.171 \\
\midrule
\multirow{2}{*}{Interstitial} & Formation Energy (eV) & 3.300 & 1.540 & 0.820 \\
& Zero-Phonon Line (eV) & 0.279 & 0.223 & 0.235 \\
\bottomrule
\end{tabular}
\caption{Projection scores corresponding to the best performing combination of scaling and descriptor for the ridge (R), the kernel ridge (KR), and the multilayer perceptron (MLP) model for formation energy and zero-phonon line predictions in vacancies, substitutions, and interstitials. These projected scores correspond to the best scores of each model highlighted in Table~\ref{tab:all_results}.}
\label{tab:projections}
\end{table}

{%
\captionsetup[subfigure]{labelformat=empty}

\begin{figure}
    \centering

    \makebox[\linewidth][c]{%
        \raisebox{-0.5\height}{\rotatebox{90}{\normalsize Average MAE (eV)}}%
        \hspace{2mm}
        \begin{minipage}{0.9\linewidth}
            \centering

            \hspace{9mm}
            \begin{subfigure}[b]{0.8\linewidth}
                \centering
                \includegraphics[width=\linewidth]{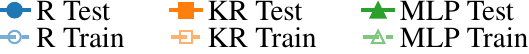}
            \end{subfigure}

            \begin{subfigure}[b]{\linewidth}
                \begin{overpic}[width=\linewidth]{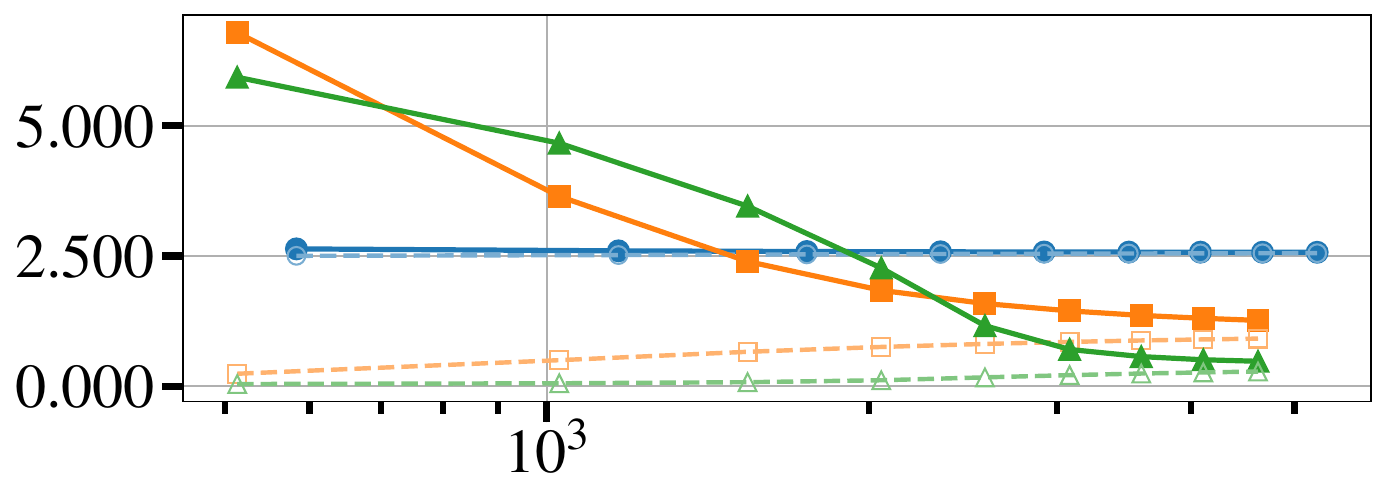}
                    \put(15,12.5){\small (a)}
                \end{overpic}
                \caption{}\label{fig:mae_data_points_train_a}
            \end{subfigure}

            \vspace{-8.2mm}
            \begin{subfigure}[b]{\linewidth}
                \begin{overpic}[width=\linewidth]{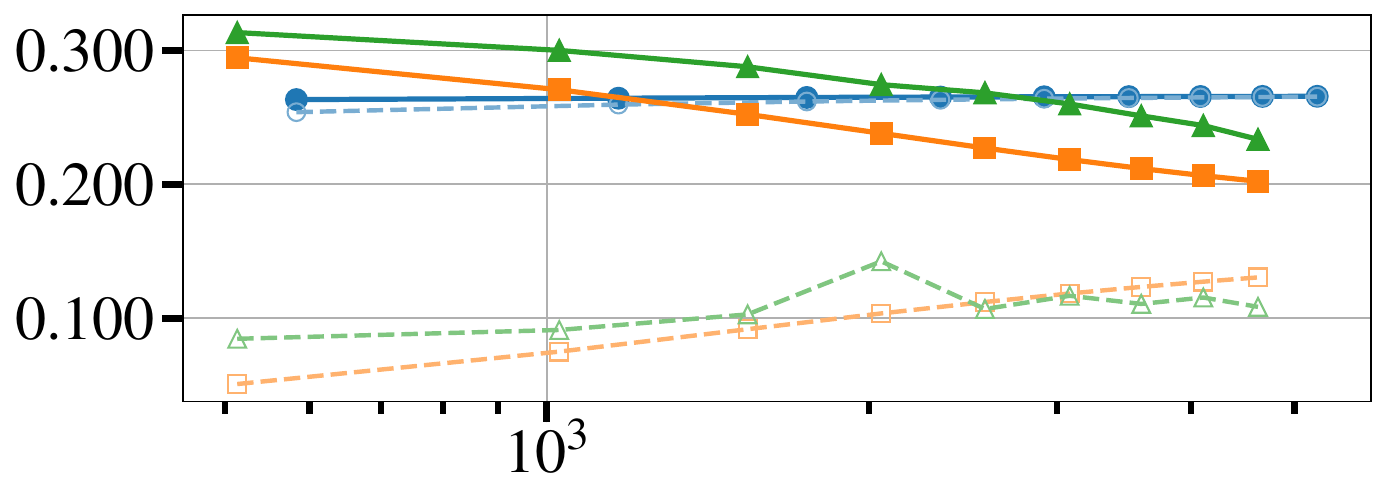}
                    \put(15,17){\small (b)}
                \end{overpic}
                \caption{}\label{fig:mae_data_points_train_b}
            \end{subfigure}

            \vspace{-7.1mm}

            \begin{subfigure}[b]{\linewidth}
                \begin{overpic}[width=\linewidth]{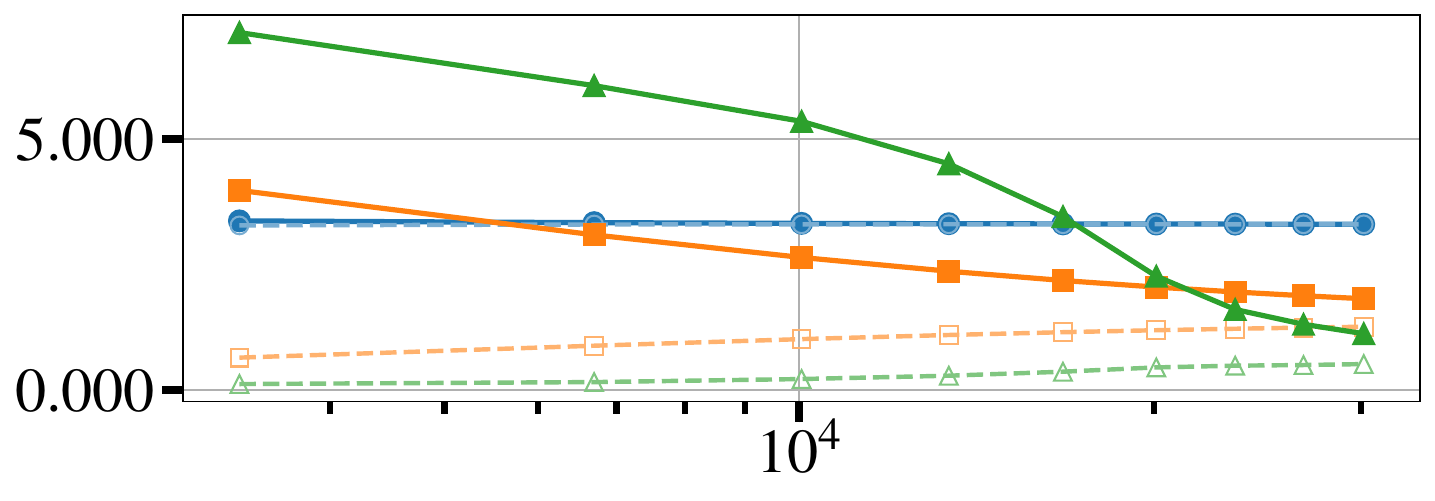}
                    \put(15,14){\small (c)}
                \end{overpic}
                \caption{}\label{fig:mae_data_points_train_c}
            \end{subfigure}

            \vspace{-7.1mm}

            \begin{subfigure}[b]{\linewidth}
                \begin{overpic}[width=\linewidth]{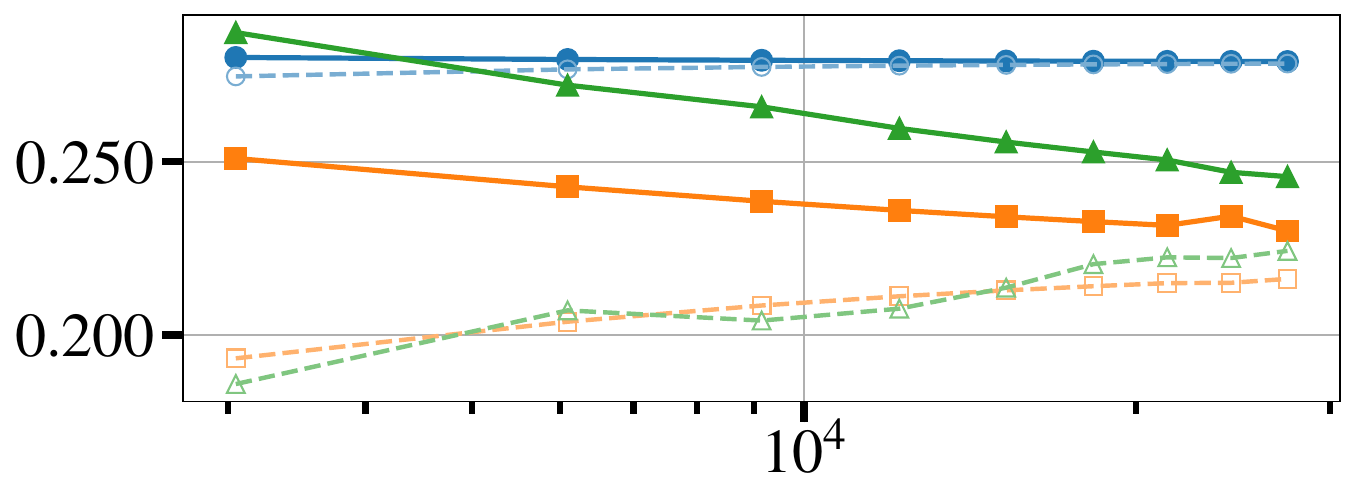}
                    \put(15,13){\small (d)}
                \end{overpic}
                \caption{}\label{fig:mae_data_points_train_d}
            \end{subfigure}

            \vspace{-7mm}
            \centering
            \normalsize Training data points (Log10 scale)
        \end{minipage}
    }
    \caption{Learning curves of the average MAE using split ratio 10-fold cross-validation for train and test data as a function of the average number of training data points for the ridge (R), the kernel ridge (KR), and the multilayer perceptron (MLP) model, where each uses its respective best combination of descriptor, scaling and hyperparameters given in Table~\ref{tab:all_results}. (a) and (b) show formation energy and zero-phonon line predictions for vacancies and substitutions, respectively. Similarly, (c) and (d) show respectively the same thing but for interstitials.}
    \label{fig:mae_data_points_train}
\end{figure}
}

Scatter plots of predicted vs.~true values can be used to evaluate model precision. In Fig.~\ref{fig:scatter_plots} predictions from one of the fold combinations in the 10-fold cross-validation averages are shown for the best scores highlighted in Table~\ref{tab:all_scores}.

{%
\captionsetup[subfigure]{labelformat=empty}

\begin{figure}
    \centering

\makebox[\linewidth][c]{%
    \raisebox{-0.5\height}{\rotatebox{90}{\normalsize ML model predicted energy (eV)}}%
    \hspace{1mm}
    \begin{minipage}{0.9\linewidth}
        \centering

        \begin{subfigure}[b]{1\linewidth}
            \centering
            \includegraphics[width=\linewidth]{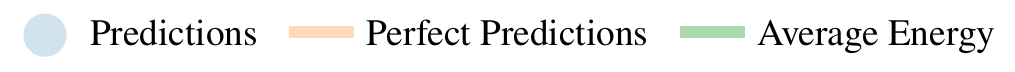}
        \end{subfigure}

        \vspace{-1.25mm}

        \begin{subfigure}[b]{0.49\linewidth}
            \centering
            \begin{overpic}[width=\linewidth,keepaspectratio,
                trim=0 0 0 50mm,clip]{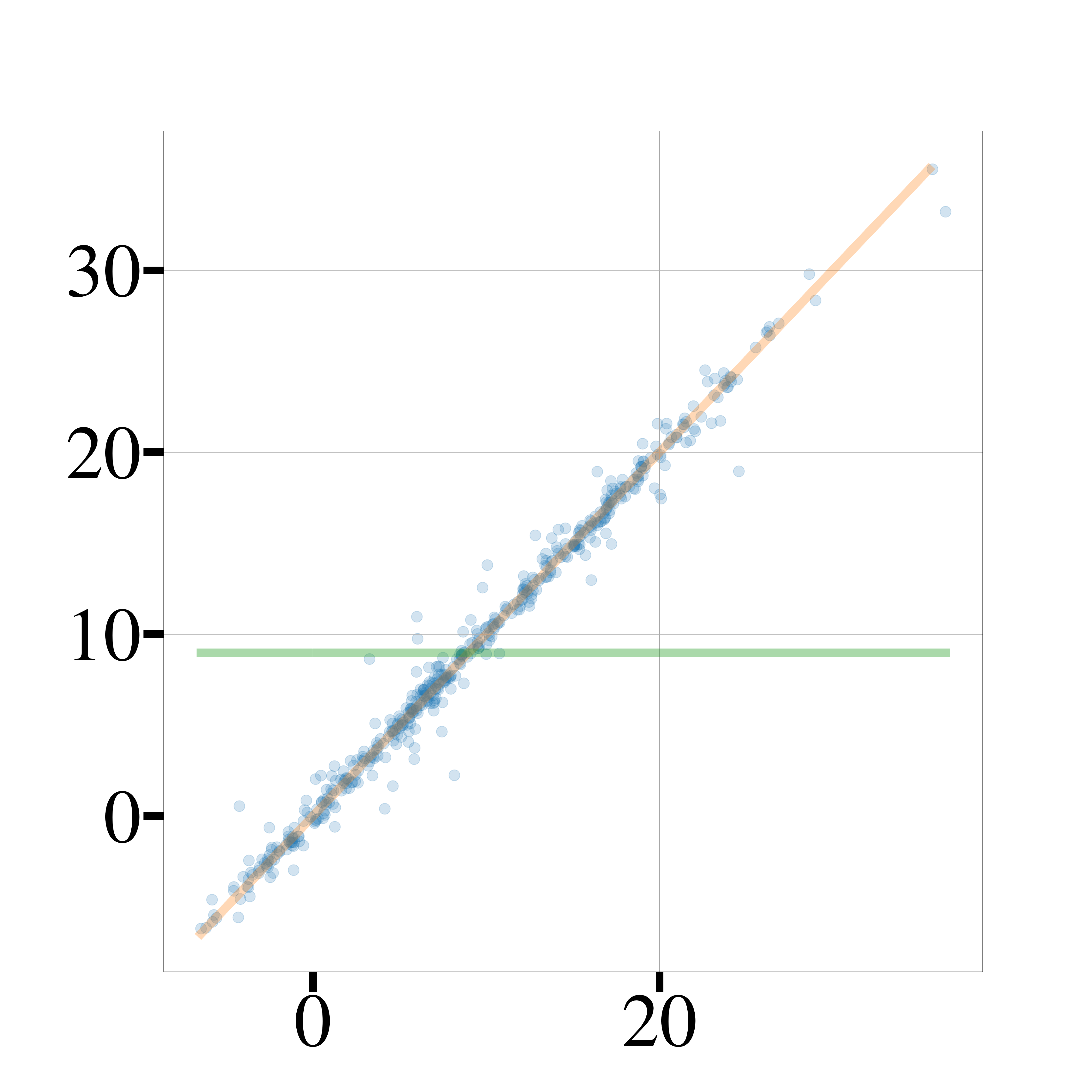}
                \put(70,16){\small (a)}
            \end{overpic}
            \caption{}
            \label{fig:scatter_a}
        \end{subfigure}
        \hfill
        \begin{subfigure}[b]{0.49\linewidth}
            \centering
            \begin{overpic}[width=\linewidth,keepaspectratio,
                trim=0 0 0 50mm,clip]{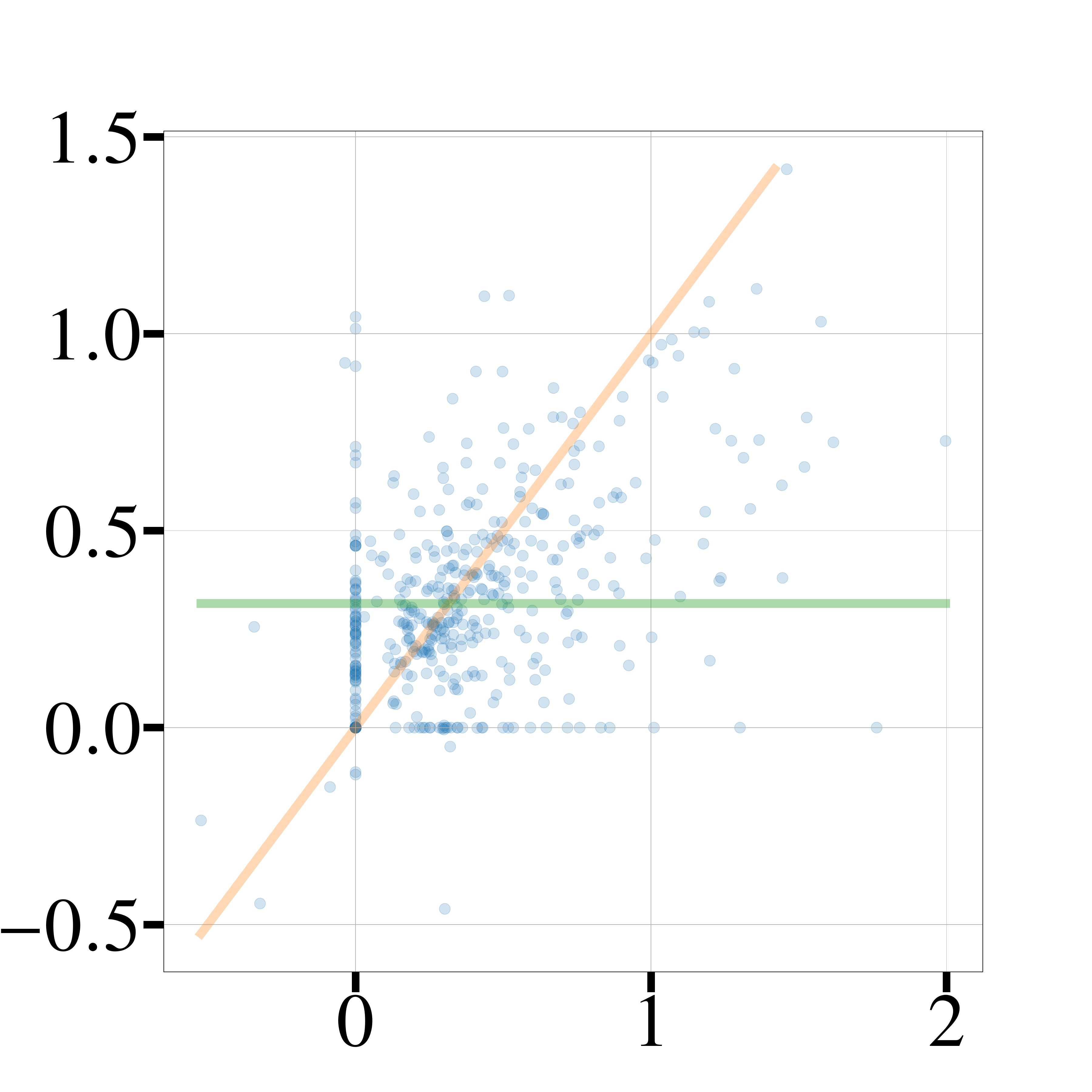}
                \put(69,21){\small (b)}
            \end{overpic}
            \caption{}
            \label{fig:scatter_b}
        \end{subfigure}

        \vspace{-8mm}

        \begin{subfigure}[b]{0.49\linewidth}
            \centering
            \begin{overpic}[width=\linewidth,
                trim=0 0 0 50mm,clip]{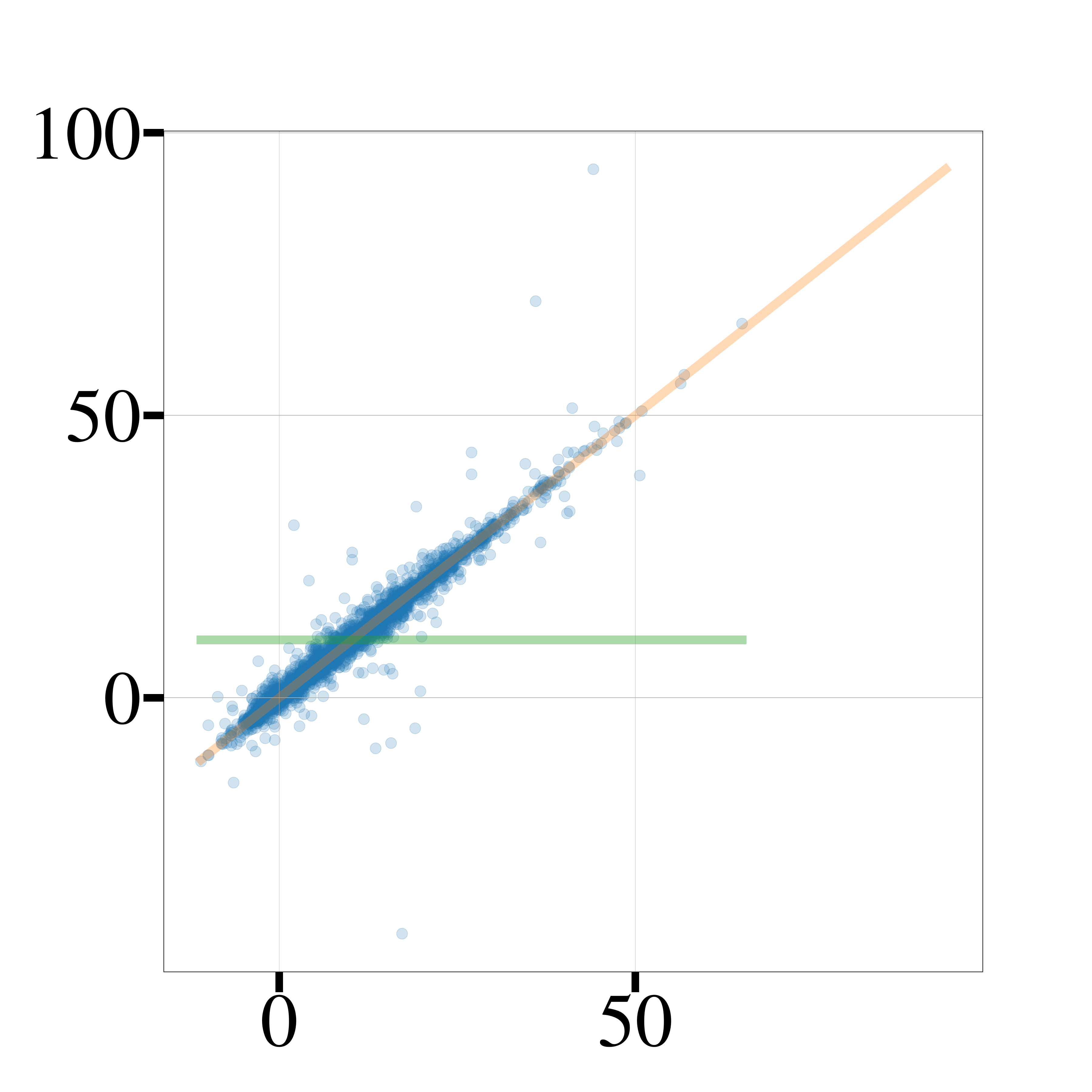}
                \put(77,22){\small (c)}
            \end{overpic}
            \caption{}
            \label{fig:scatter_c}
        \end{subfigure}
        \hfill
        \begin{subfigure}[b]{0.49\linewidth}
            \centering
            \begin{overpic}[width=\linewidth,
                trim=0 0 0 50mm,clip]{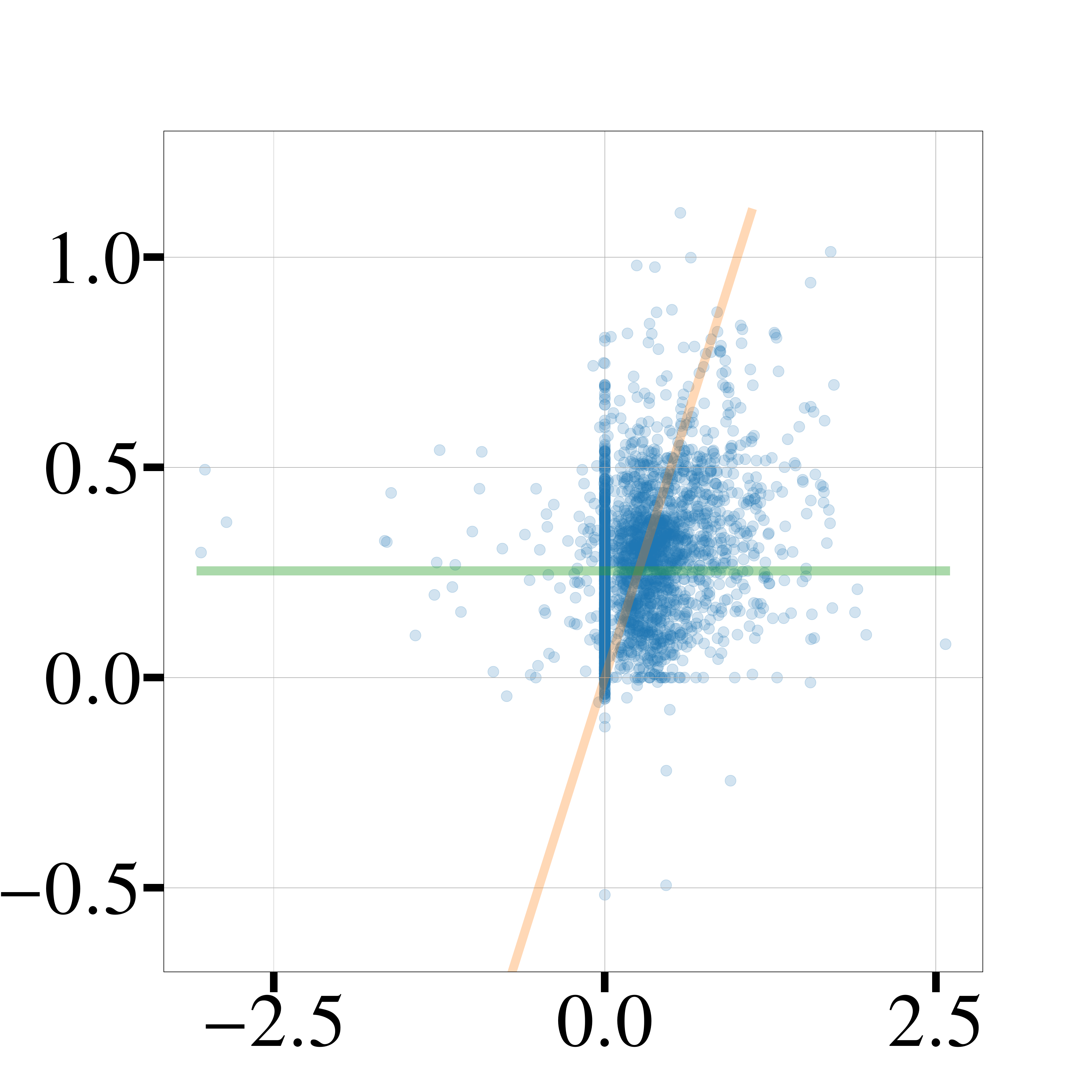}
                \put(68,22){\small (d)}
            \end{overpic}
            \caption{}
            \label{fig:scatter_d}
        \end{subfigure}

        \vspace{-6mm}
        \centering
        \normalsize DFT energy (eV)
    \end{minipage}
}
    \caption{Scatter plots of the predicted energy as a function of the true energy. (a) and (b) show respectively formation energy and zero-phonon line predictions for vacancies and substitutions. Similarly, (c) and (d) show respectively the same thing but for interstitials. These predictions each correspond to one out of the 10 mean absolute error scores used to calculate the final best score for each defect type and predicted energy encircled in red in Table~\ref{tab:all_scores}.}
    \label{fig:scatter_plots}
\end{figure}
}

\section{Discussion}\label{sec:discussion}

As shown in Table~\ref{tab:all_scores}, the smallest average errors of the formation energy and zero-phonon predictions are around \(0.5\) eV and \(0.2\) eV for vacancies and substitutions, respectively. This result matches the screening accuracy of high-throughput calculations, which allow for ML to sort out uninteresting point defects. The respective errors for interstitials are around \(1.1\) and \(0.23\) eV, which is worse but still enough to be used as a coarse-grained filter. The previous results show that ML can be used to alleviate the computational cost of high-throughput calculations.

The Atomic Descriptor gives the lowest reported MAEs for formation energies and ZPLs in vacancies and substitutions, but is effectively tied with the Wyckoff Coordinate Descriptor for formation energies (0.478 vs.~0.485~eV). The Atomic Descriptor is also among the best-performing descriptors for interstitial ZPLs (Table~\ref{tab:all_scores}). The Wyckoff descriptors retain the atomic-number information but distribute it among separate vector entries according to site category. This provides additional structural information while changing how similarities between defect systems are represented. One possible explanation for their poorer performance on some tasks is that this separation makes relationships shared across site categories harder to learn from a finite dataset. The results are consistent with this explanation but do not establish it, since the comparisons also involve differences in descriptor dimensionality and optimized model configurations. Nevertheless, the comparisons restricted to the same retained samples confirm that the additional site encoding does not improve the best vacancy and substitution predictions obtained here.

The matched-dataset comparisons (the ``Compare'' entries in Table~\ref{tab:all_results}) show that the advantage of the Wyckoff descriptors for interstitial formation-energy prediction persists when the same defect configurations are used. For the standardized multilayer perceptron, the MAE decreases from 2.582 eV with the Atomic Descriptor to 1.633 eV with the Wyckoff Letter Compare Descriptor and 1.182 eV with the Wyckoff Coordinate Compare Descriptor. The corresponding MAEs for kernel ridge regression using unscaled data are 2.530, 2.410, and 1.841 eV, respectively. Thus, relative to the Atomic Descriptor, the Wyckoff Coordinate Compare Descriptor reduces the MAE by approximately 54\% for the multilayer perceptron and 27\% for kernel ridge regression. Hence, the improved performance obtained with the full Wyckoff datasets is not explained solely by the additional configurations they retain.

The formation-energy learning curves show that the relative performance of the nonlinear models depends on the amount of training data. Kernel ridge regression can outperform the MLP at intermediate training sizes, whereas the MLP achieves lower validation MAEs at the largest training sizes considered. This suggests that the tested MLP configurations make more effective use of larger datasets for these tasks. Their substantial improvement over linear ridge regression also supports the importance of nonlinear relationships among the descriptor features.
However, because kernel ridge regression also represents nonlinear relationships, these comparisons do not establish that greater model flexibility alone explains the MLP’s advantage.

Kernel ridge regression achieves lower ZPL MAEs than the multilayer perceptron for both defect classes (Table~\ref{tab:all_results}). Both model classes can represent nonlinear relationships, and their relative predictive performance depends on the descriptors, training data, regularization, and optimization. The present results therefore establish a performance advantage for the tested kernel ridge configurations, without identifying a single explanation for that advantage.
The different MLP architectures and solvers selected for formation-energy and ZPL prediction indicate that the preferred model configurations depend on the target property. These differences do not, by themselves, establish distinct underlying physical relationships. Similarly, the fluctuations in the MLP training MAEs in Figs. 3b and 3d suggest that sensitivity to initialization and optimization warrants investigation, but the learning curves alone cannot determine whether incomplete optimization accounts for the higher ZPL errors.
The scatter plots in Figs.~\ref{fig:scatter_b}~and~\ref{fig:scatter_d} also show that substantial discrepancies remain between predicted and reference ZPLs even for the best-performing models. More informative descriptors or changes to the training procedure may improve these predictions. The current comparisons do not determine which changes would be most effective.

The scatter plots in Fig.~\ref{fig:scatter_plots} shows that the performance for formation energy predictions are better than for ZPLs, which suggest that the latter is more difficult to predict from the provided descriptors. Furthermore, Table~\ref{tab:all_scores} shows the best MLP MAEs for ZPL prediction are similar for vacancy-substitution and interstitial defects, 0.234 and 0.246 eV respectively, whereas the corresponding best formation-energy MAEs differ more substantially, 0.478 and 1.123 eV. Hence, the absolute ZPL prediction error varies less between defect classes for these separately trained models. Apparently the additional site information is useful to substantially improve the interstitial formation-energy predictions, but the same does not hold for the best ZPL predictions. It is possible that including other low-cost features more directly related to the defect electronic structure in the descriptors could be more effective at improving the ZPL predictions.

A relevant future extension of this work may be to train models directly on defect-pair binding energies, defined relative to the sum of the formation energies of the isolated constituent defects. Since the single-defect configuration space can be characterized much more exhaustively, its calculated formation energies could provide a reference, leaving the model to learn the interaction contribution associated with forming a double defect. It is possible such an approach could improve prediction accuracy or data efficiency compared with the approach of this work of directly learning double-defect formation energies.

The case where the Wyckoff Letter Descriptor and the Wyckoff Coordinate Descriptor outclass the Atomic Descriptor is for interstitial formation energy predictions, as seen in Table~\ref{tab:all_scores}. Between the two, the Wyckoff Coordinate Descriptor achieves the best performance, suggesting that more information about the neighboring atomic structure symmetry is beneficial in this case. A previous study also reported improved interstitial formation-energy predictions in diamond when information about the local atomic environment was included \cite{paul_master}.

\section{Conclusions}\label{sec:conclusion}

For double defects containing vacancies and substitutions, the best MAEs are $0.478\ \mathrm{eV}$ for formation energies and \(0.202\) eV for zero-phonon lines. For interstitials, the corresponding values are \(1.123\) eV  and \(0.230\) eV. These predictions are precise enough for the ML models presented in this work to be immediately useful as a prescreening filter of enumerated interstitials in high-throughput studies. Our approach also remains promising for further refinement, which may be able to bring down errors to a level where, for some applications, costly DFT calculations of large defect systems can be entirely avoided.

The best performance in three of the four prediction tasks is obtained by a descriptor that consists of: an atomic number representation of the point defects, the distance between each unique pair of point defects, and the charge and spin state of the defect cluster. The exception is the interstitial formation energy, where the additional site information provided by the Wyckoff Coordinate Descriptor substantially improves performance.  The Wyckoff Letter Descriptor also improves predictions for this task, but gives higher MAEs than the Wyckoff Coordinate Descriptor.

The scatter plots in Fig.~\ref{fig:scatter_plots} show substantial discrepancies remain between predicted and reference ZPLs. A future direction is to investigate whether additional descriptor information related to the defect electronic structure could improve these predictions.

\section{Data availability}

The data used to train the machine learning models in this paper are publicly available in the ADAQ database (\url{https://defects.anyterial.se}). Our machine learning scripts, trained models, and the outputs from applying those models as presented in tables and figures are available in a source code repository at \url{https://si.anyterial.se/mldefects}.

\section{Conflicts of interest}

There are no conflicts of interest to declare.

\section*{Acknowledgments}

RA acknowledges support from the Swedish Government Strategic Research Area Swedish e-science Research Centre (SeRC).
JD acknowledges support from the Swedish Research Council (VR) Grant No. 2022-00276, Zenith Career Grant no. 26.15 from Linköping University, and the Magnus Bergvall Foundation (Grant no. 2025-602).
The computations were enabled by resources provided by the National Academic Infrastructure for Supercomputing in Sweden (NAISS), partially funded by the Swedish Research Council through grant agreement no. 2022-06725.

\bibliography{references}

\end{document}